\documentclass{llncs}

\usepackage{graphicx}%
\usepackage{multirow}%
\usepackage{amsmath,amssymb,amsfonts}%
\usepackage{mathrsfs}%
\usepackage[title]{appendix}%
\usepackage{xcolor}%
\usepackage{textcomp}%
\usepackage{manyfoot}%
\usepackage{booktabs}%
\usepackage{algorithm}%
\usepackage{algorithmicx}%
\usepackage{algpseudocode}%
\usepackage{listings}%
\usepackage{float}
\usepackage{caption}
\usepackage{tikz}
\usepackage{placeins}
\usepackage{bbding} 
\usepackage{cite}
\usepackage{etoolbox}
\usepackage[T1]{fontenc}
\usepackage{lmodern}

\usepackage[
  a4paper,          
  textwidth=15cm,   
  textheight=24cm,  
  heightrounded,    
  hratio=1:1,       
  vratio=2:3        
]{geometry}

\usetikzlibrary{positioning,calc,arrows.meta}

\algblock{Start}{End}

\algblockdefx[NAME]{START}{END}%
[2][Unknown]{Start #1(#2)}%
{Ending}

\algblockdefx[NAME]{}{OTHEREND}[1]{Until (#1)}

\spnewtheorem{construction}[theorem]{Construction}{\bfseries}{\rmfamily}
\AtBeginEnvironment{proof}{\setcounter{case}{0}}
\title{On cryptographically non-linear all-or-nothing transforms\thanks{This work was supported by the Natural Sciences and Engineering
Research Council of Canada (NSERC) through Discovery Grant RGPIN-2024-06660 and
Discovery Launch Supplement DGECR-2024-00001, and by start-up funding from
Toronto Metropolitan University. A.\ Devaraj was supported by an Undergraduate
Research Opportunities (URO) award from the Faculty of Science, Toronto
Metropolitan University.}}
 \author{Anantajeet Devaraj and Navid Nasr Esfahani~\Envelope}
\institute{{Department of Computer Science}\\
Toronto Metropolitan University\\ 
Toronto, Ontario,M5B 2K3, Canada\\ \email{nne@torontomu.ca}}

\begin{document}

\maketitle

\abstract{An unconditionally secure $(t, s, v)$-all-or-nothing transform (AONT) is a bijection from $s$ input elements to $s$ output elements over an alphabet of size $v$, such that given any $s - t$ output elements, no information can be obtained about any $t$ input elements. While linear AONTs have been extensively studied, comparatively little is known about the existence of non-linear AONTs. In this paper, we investigate unconditionally secure cryptographically non-linear  $(t, s, v)$-AONTs, distinguishing between non-additivity, non-affinity, and cryptographic non-linearity. To achieve this, we first present a computational method for enumerating general AONTs focusing on the case $(t, 3, 3)$-AONTs for $t = 1 ~{\rm  and }~ 2$, establishing the existence of non-additive AONTs. We then introduce a general construction of non-additive and non-affine AONTs from linear ones using coordinate-wise symbol permutations. We measure the non-linearity of the AONT instances generated from this construction, using differential $\Delta_F$-uniformity and Walsh magnitude. Finally, we establish theoretical conditions under which both these measures are reduced, and present statistical results examining the non-linearity of AONTs obtained through our construction.}


\section{Introduction}\label{sec1}

All-or-nothing transforms (AONTs) were proposed in 1997 by Rivest \cite{rivest-AONT} in the context of block ciphers as a mode of operation, focusing on AONTs that are computationally secure. Unconditionally secure AONTs were introduced and studied by Stinson \cite{stinson-AONT}. Recent works have expanded the theory of AONTs by developing broader frameworks and exploring new classes of unconditionally secure AONTs \cite{des-allornothing,egs-results,es-secuirtyproperities,es-rectangular,gaems-itsofaonts,es-asym,wcj-2ppexist}. We begin with some important definitions. 

\begin{definition}
Let $\Sigma$ be an alphabet of size $v$, let $s$ be a positive integer, and let $\phi$ be a mapping
from $\Sigma^s$ to $\Sigma^s$, i.e.,
$(y_1, y_2, \ldots, y_s) = \phi(x_1, x_2, \ldots, x_s)$
for $(x_1, x_2, \ldots, x_s) \in \Sigma^s$ and
$(y_1, y_2, \ldots, y_s) \in \Sigma^s$.
Then $\phi$ is an unconditionally secure $(t,s,v)$-AONT if it satisfies the following conditions:
\begin{enumerate}
  \item $\phi$ is a bijection.
  \item If any $s - t$ of the output elements are fixed, then any $t$ input element can take on all possible values in $\Sigma^t$.
\end{enumerate}
\end{definition}

\begin{definition}
Let $\mathbb{F}_q$ be a finite field of order $q$. An AONT with alphabet
$\mathbb{F}_q$ is \emph{linear} if each $y_i$ is an $\mathbb{F}_q$-linear function
of $x_1, \ldots, x_s$. Then we can write
\begin{equation}
\mathbf{y} = \phi(\mathbf{x}) = \mathbf{x} M^{-1}
\quad \text{and} \quad
\mathbf{x} = \phi^{-1}(\mathbf{y}) = \mathbf{y} M,
\end{equation}
where $M$ is an invertible $s \times s$ matrix with entries from $\mathbb{F}_q$.
Subsequently, when we refer to a ``linear AONT'', we mean the matrix $M$ that
transforms $\mathbf{y}$ to $\mathbf{x}$, as specified above.
\end{definition}

 For $M$ to define a valid $(t,s,q)$-AONT, every $t\times t$ submatrix of $M$  obtained by deleting rows and columns must be invertible \cite{des-allornothing}.

\begin{definition}[Orthogonal Array]
Let $\mathcal{X}$ be an alphabet of size $v$, and let $A$ be an $N \times k$ array with
entries from $\mathcal{X}$. We call $A$ an $(N,k,v)$-array. Label the columns of $A$
by the set $C=\{1,\ldots,k\}$. For any subset $D \subseteq C$, let $A_D$ denote the
subarray obtained from $A$ by retaining only the columns indexed by $D$. We say that
$A$ is \emph{unbiased with respect to $D$} if, among the rows of $A_D$, every
$|D|$-tuple over $\mathcal{X}$ occurs exactly $N / v^{|D|}$ times.

An \emph{orthogonal array}, denoted by $\mathrm{OA}_{\lambda}(s,k,v)$, is a
$(\lambda v^{s},k,v)$-array that is unbiased with respect to every subset
$D \subseteq C$ of size $s$. Equivalently, in any selection of $s$ columns, every
$s$-tuple over $\mathcal{X}$ appears exactly $\lambda$ times. By convention, the
subscript $\lambda$ is omitted when $\lambda = 1$.
\end{definition}

By definition, orthogonal arrays are isomorphic under permutations of rows, permutations of columns, and independent permutations of symbols within columns \cite{Orthogonal-array}.

The following theorem by D'Arco et al.\ \cite{des-allornothing} formulates the equivalency of $t$-AONTs to certain classes of orthogonal arrays. It is a generalized version of a similar theorem by  Stinson \cite{stinson-AONT}.
\begin{theorem}\cite[Theorem 34]{des-allornothing} \label{Thrm:array-equivalence}
A $(t,s,v)$-AONT is equivalent to a $(v^{s},2s,v)$-array that is
unbiased with respect to the following subsets of columns:
\begin{enumerate}
  \item $\{1,\ldots,s\}$,
  \item $\{s+1,\ldots,2s\}$, and
  \item $I \cup \{s+1,\ldots,2s\} \setminus J$, for all
        $I \subseteq \{1,\ldots,s\}$ with $|I| = t$ and all
        $J \subseteq \{s+1,\ldots,2s\}$ with $|J| = t$.
\end{enumerate}
\end{theorem}

Theorem \ref{Thrm:array-equivalence} characterizes a $(t,s,v)$-AONT in terms of unbiased arrays, implying that for any $(t,s,v)$-AONT, selecting any $t$ input columns together with any $s-t$ output columns yields an orthogonal array in which every $s$-tuple occurs exactly once.

\begin{corollary}\label{Cor:AONT-Inverse}
A mapping $\phi : \mathcal{X}^s \to \mathcal{X}^s$ is a $(t,s,v)$-AONT if and only if
$\phi^{-1}$ is an $(s-t,s,v)$-AONT.
\end{corollary}

In this paper, $d(\mathbf{x},\mathbf{y})$ denotes the Hamming distance between the vectors $\mathbf{x}$ and $\mathbf{y}$, and for a nonnegative integer $r$ the Hamming sphere of radius $r$ centered at $\mathbf{x}$ is the set $S_r(\mathbf{x}) = \{\, \mathbf{y} \in \mathcal{X}^s : d(\mathbf{x},\mathbf{y}) \le r \,\}$.

To determine the level of the non-linearity for various non-linear AONT constructions, we need to use measures of non-linearity. There are many such measures  used in cryptography \cite{bk-pnfandcrypto}, in the context of differential cryptanalysis. In this paper, we will use differential $\Delta_F$-uniformity  and Walsh transform \cite{bk-pnfandcrypto} as our non-linearity measures. We use the following definitions by Blondeau and Nyberg \cite{bk-pnfandcrypto}.

\begin{definition}[Differential $\Delta_F$-uniformity]
Let $A$ and $B$ be finite Abelian groups and let
\[
F: A \to B
\]
be a function.

Given $a \in A$ and $b \in B$, the equation
\[
F(x+a)-F(x)=b
\]
for $x \in A$ is called a differential of $F$ with input difference $a$ and output difference $b$.

Define
\[
\delta(a,b)
=
\left|
\left\{
x \in A :
F(x+a)-F(x)=b
\right\}
\right|.
\]

We denote by $\Delta_F$ the positive integer
\[
\Delta_F
=
\max_{\substack{a \in A,\ a \neq 0 \\ b \in B}}
\delta(a,b).
\]

Then $F$ is said to be $\Delta_F$-uniform.
\end{definition}

\begin{definition}[Walsh Transform and Walsh Magnitude]
Let
\[
F : \mathbb{F}_q^s \to \mathbb{F}_q^s
\]
be a function, where $q=p^m$ and $p$ is prime.

The Walsh transform of $F$ is defined by
\[
W_F(a,b)
=
\sum_{x \in (\mathbb{F}_q)^s}
\omega^{\operatorname{Tr}(\langle b,F(x)\rangle-\langle a,x\rangle)},
\]
where
\[
\omega=e^{2\pi i/p},
\]
and
\[
\operatorname{Tr}(x)
=
x+x^p+x^{p^2}+\cdots+x^{p^{m-1}}
\]
is the absolute trace map from $\mathbb{F}_q$ to $\mathbb{F}_p$.

The maximum Walsh magnitude of $F$ is defined as
\[
W_{\max}(F)
=
\max_{\substack{
a,b \in (\mathbb{F}_q)^s\\
b \neq 0
}}
|W_F(a,b)|.
\]

The Walsh transform measures the correlation between $F$ and affine functions, with larger $W_{\max}$ indicating stronger linear structure.
\end{definition}

Next we present a universal lower bound on $W_{max}$ derived from Parseval's identity.
\begin{lemma}[Parseval's Identity\cite{carlet-boolean}] \label{lem:Parseval}
Let $F:\mathbb{F}_q^s\rightarrow\mathbb{F}_q^s$ be a mapping. For any nonzero output mask $b\in\mathbb{F}_q^s$, the Walsh coefficients satisfy
\[
\sum_{a\in\mathbb{F}_q^s}|W_F(a,b)|^2=q^{2s}.
\]

By Parseval's identity, the maximum Walsh magnitude satisfies
\[
q^{s/2}\leq W_{\max}\leq q^s
\]
The upper bound $q^s$ is achieved by linear mappings.
\end{lemma}

Next we define the three types of non-linearity for a $(t,s,q)$-AONT to be non-linear, as we will use in this paper, namely: \emph{non-additivity, non-affinity, and cryptographic non-linearity}.
\begin{definition}[Non-additive AONT]
Let $q=p^m$ be a prime power, we identify the symbols of the AONT with the elements of the finite field $\mathbb{F}_q$. A $(t,s,q)$-AONT
\[
\phi : \mathbb{F}_q^s \to \mathbb{F}_q^s
\]
is called non-additive if there exist
\[
\mathbf{x}, \mathbf{y} \in \mathbb{F}_q^s
\]
such that
\[
\phi(\mathbf{x}) + \phi(\mathbf{y})
\neq
\phi(\mathbf{x}+\mathbf{y}).
\]

In this case, $\phi$ cannot be represented by multiplication by a matrix over $\mathbb{F}_q$.
\end{definition}

It should be noted that affine maps, i.e., $xM+c$, trivially satisfy the definition above.

\begin{definition}[Non-affine AONT]
Let $q=p^m$ be a prime power, we identify the symbols of the AONT with the elements of the finite field $\mathbb{F}_q$. A $(t,s,q)$-AONT
\[
\phi : \mathbb{F}_q^s \to \mathbb{F}_q^s
\]
is called non-affine if for any fixed $\mathbf{c} \in \mathbb{F}_q^s$, there exist
\[
\mathbf{x}, \mathbf{y} \in \mathbb{F}_q^s
\]
such that
\[
\phi(\mathbf{x}) + \phi(\mathbf{y})
\neq
\phi(\mathbf{x}+\mathbf{y})+\mathbf{c}
\]

In this case, $\phi$ cannot be represented as multiplication by a matrix over $\mathbb{F}_q$ translated by a constant.
\end{definition}

\begin{definition}[Cryptographically non-linear AONT]
Let $q=p^m$ be a prime power, we identify the symbols of the AONT with the elements of the finite field $\mathbb{F}_q$. A $(t,s,q)$-AONT
\[
\phi : \mathbb{F}_q^s \to \mathbb{F}_q^s
\]
is called cryptographically non-linear if 
\[
\Delta_{\phi} < q^s
\]
and
\[
W_{max}(\phi) < q^s
\]
\begin{remark}
If $\Phi_x$ is the set of all AONTs with $x$ property, then 
\[\Phi_{\rm cryptographically\ non-linear}\subset \Phi_{\rm non-affine}\subset \Phi_{\rm non-additive}\]
\end{remark}

\end{definition}
We also fix the following notational convention used throughout the paper. For a row $\mathbf{v} = (v_1, \ldots, v_{2s})$ of the orthogonal array representation of an AONT, we write $\mathbf{v}_{[1,s]}$ and $\mathbf{v}_{[s+1,2s]}$ to refer to its input portion $(v_1, \ldots, v_s)$ and its output portion $(v_{s+1}, \ldots, v_{2s})$, respectively.

\subsection{Organization of the Paper}\label{subsec2}
Section \ref{sec:relwork} offers a brief summary of previous related work. Section \ref{sec:new-theoretical} presents two corollaries on general AONTs derived from classic results. Section \ref{sec:enum-all-general-t33} focuses on a computational method to enumerate all general $(t,3,3)$-AONTs searching for non-linear transforms. In Section \ref{sec:const-non-linear}, the construction of non-additive and non-affine AONTs from linear ones is presented. Section \ref{sec:measuring-non-linearity} examines the non-linearity of AONTs generated using the construction from Section \ref{sec:const-non-linear}, establishing conditions under which 
$\Delta_{F}$-uniformity and $W_{max}$ can be reduced, and deriving simple bounds on the reductions achievable. In Section \ref{sec:Randomized-Search} we present computational results for larger AONT parameters and investigate the average degree of non-linearity exhibited by our constructions. Finally, Section \ref{sec:conclusion} concludes the paper.

\section{Related Work}\label{sec:relwork}
This section reviews prior work in two areas relevant to our results: the theory of all-or-nothing transforms and measures of non-linearity in cryptography.

\subsection{All-or-Nothing Transforms}
Rivest \cite{rivest-AONT} introduced AONTs in 1997 as a mode of operation for block ciphers, designed to impede brute-force attacks when the key length cannot be increased. Rivest's original construction was computationally secure and relied on a random key. Stinson \cite{stinson-AONT} formalized unconditionally secure AONTs and established their equivalence to unbiased arrays, opening the door for a purely combinatorial treatment of the problem. Building on this foundation, subsequent work has explored several directions.

D'Arco et al.\ \cite{des-allornothing} studied the existence of $(t,s,v)$-AONTs, with an emphasis on the case of $t =v=2$, while providing structurally close constructions and non-existence results. Esfahani et al.\ \cite{egs-results} extended the results to arbitrary alphabet sizes. Wang et al.\ \cite{wcj-2ppexist} established the existence of linear $(2,p,p)$-AONTs for all prime values of $p$. Esfahani and Stinson introduced asymmetric AONTs in which any $t$ input blocks remain unrevealed in the absence of $t'>t$ output blocks \cite{es-asym} and proposed rectangular, range, and restricted variants of AONTs \cite{es-rectangular}. They also investigated the security properties of AONTs beyond perfect security \cite{es-secuirtyproperities}. Most recently, Gu et al.\ \cite{gaems-itsofaonts} further studied the information-theoretic security of unconditionally secure AONTs.

The constructions appearing in the literature above are all linear. The matrix representation reduces the unbiasedness conditions in Theorem \ref{Thrm:array-equivalence} to the algebraic constraints on the matrix $M$.

\subsection{Non-linearity in Cryptography}
Non-linearity is a central design criterion for symmetric primitives such as S-boxes and Boolean functions, where resistance to differential \cite{nyberg-apn} and linear \cite{matsui-linear} cryptanalysis depends on the function deviating substantially from affine behaviour. Two standard quantitative measures of non-linearity are the differential $\Delta_F$-uniformity \cite{nyberg-apn} and the maximum Walsh magnitude $W_{\max}$ \cite{matsui-linear}. A function with low $\Delta_F$-uniformity offers strong resistance to differential cryptanalysis, while small $W_{\max}$ indicates weak correlation with affine functions and hence resistance to linear cryptanalysis. We refer to the survey of Blondeau and Nyberg \cite{bk-pnfandcrypto} and the book by Carlet \cite{carlet-boolean} for a unified presentation of these measures and of the classes of functions that achieve their extremal values, including perfect non-linear (PN) functions, almost perfect non-linear (APN) functions \cite{nyberg-apn}, and bent functions \cite{rothaus-bent}.

While these measures have been extensively studied for vectorial Boolean functions \cite{carlet-boolean}, in particular for resilient functions \cite{sm-nlbounds-RF,tarannikov, zz-RF}, they have not previously been applied to AONTs. Esfahani et al.\ \cite{egs-results} showed that any $(t,s,v)$-AONT can be used to construct an $(s, s-t, t,v)$-resilient function. We start with non-additive AONTs, 
and end with a search for cryptographically non-linear AONTs, in the sense of having small $\Delta_F$-uniformity and small Walsh magnitude. 

\section{Theoretical Results for Search and Construction}\label{sec:new-theoretical}
In this section, we will introduce two corollaries which will be used in searching for non-linear AONTs and for constructing non-linear AONTs from linear ones, in future sections. These corollaries are based on well-known classical results.

The corollary below is a direct result of Theorem \ref{Thrm:array-equivalence}, and it presents a lower bound on Hamming distance between the second half of the rows of the orthogonal array representation of an AONT. This result will be used to optimize the exhaustive search for non-linear AONTs. It is a consequence of the relationship between the strength of an orthogonal array and the minimum distance between its rows \cite{bush1952, hss-oa1999}.
\begin{corollary}
\label{Cor:distance}
Suppose $\phi : \mathcal{X}^s \to \mathcal{X}^s$ is a $(t,s,v)$-AONT. Let $\mathbf{a}, \mathbf{c}\in \mathcal{X}^s$ where $\mathbf{a}\neq \mathbf{c}$, $\phi(\mathbf{a}) = \mathbf{b}$, and $\phi(\mathbf{c}) = \mathbf{d}$. If $d(\mathbf{a}, \mathbf{c})\le s-t$ then $d(\mathbf{b}, \mathbf{d})\ge t+1$.
\end{corollary}
\begin{proof}
Assume for the sake of contradiction that there exists $\mathbf{a}$ and $\mathbf{c}$, $\mathbf{a}\neq \mathbf{c}$, both in $\mathcal{X}^s$ with $\phi(\mathbf{a}) = \mathbf{b}$ and $\phi(\mathbf{c})=\mathbf{d}$ such that $d(\mathbf{a}, \mathbf{c})\le s-t$ and $d(\mathbf{b}, \mathbf{d})\le t$.

By the definition of Hamming distance, $d(\mathbf{a}, \mathbf{c})\le s-t$ implies that $\mathbf{a}$ and $\mathbf{c}$ disagree in at most $s-t$ coordinates, hence agree in at least $s-(s-t) = t$ coordinates. Similarly, $d(\mathbf{b}, \mathbf{d})\le t$ implies that $\mathbf{b}$ and $\mathbf{d}$ agree in at least $s-t$ coordinates.

Let $I\subseteq \{1, \ldots, s\}$ be a set of exactly $t$ input columns on which  $\mathbf{a}$ and $\mathbf{c}$ agree, and let $J\subseteq \{s+1, \ldots, 2s\}$ be a set of exactly $s-t$ outputs on which $\mathbf{b}$ and $\mathbf{d}$ agree. By Theorem \ref{Thrm:array-equivalence}, the subarray of the array representation of $\phi$ restricted to the columns of $I\cup \{s+1, \ldots, 2s\}\setminus J$ is an orthogonal array in which every $s$-tuple over $\mathcal{X}$ occurs exactly once.

However, both the row corresponding to $\mathbf{a}$ and the row corresponding to $\mathbf{c}$ project to the same $s$-tuple on those columns. This contradicts Theorem \ref{Thrm:array-equivalence}, so $d(\mathbf{b}, \mathbf{d}) \ge t+1$. \qed
\end{proof}

The other theoretical result in this section offers a process for obtaining new AONTs from existing ones.
\begin{corollary}
\label{Cor:sym_permutation}
Suppose $\phi : \mathcal{X}^s \to \mathcal{X}^s$ is a $(t,s,v)$-AONT, and let A denote its array representation, where each row of A is of the form 
\[
 ( (x_1, ... ,x_s), \phi(x_1, ... ,x_s)) = (x_1, ... ,x_s,x_{s+1}, ... ,x_{2s})
\]
Then applying an arbitrary symbol permutation to any column of A yields a new mapping $\phi '$, which is also a $(t,s,v)$-AONT.
\end{corollary}

\begin{proof}
Let $\pi$ be a permutation of the alphabet $\mathcal{X}$, we apply $\pi$ to all entries of a column of the array $A$ and denote the resulting array by $A'$.

By construction, $A'$ is obtained from $A$ by a symbol permutation applied to a single column.
For any collection of columns specified in Theorem \ref{Thrm:array-equivalence}, the corresponding subarray of $A'$ is therefore isomorphic to the corresponding subarray of $A$. Since orthogonality is preserved under symbol permutations, each such subarray of $A'$ is an orthogonal array whenever the corresponding subarray of $A$ is.

Hence $A'$ satisfies the orthogonality conditions of Theorem \ref{Thrm:array-equivalence}, and therefore $\phi'$ is a $(t,s,v)$-AONT. \qed

\end{proof}

\section{Enumerating all general $(t,3,3)$-AONTs}\label{sec:enum-all-general-t33}

In our search for non-linear AONTs, it is natural to begin with small parameter sets, such as $(1,3,3)$- and $(2,3,3)$-AONTs. By enumerating all AONTs for these parameters, we aim to identify and better understand the structure of non-linear AONTs. We first enumerate all general $(2,3,3)$-AONTs and then apply Corollary \ref{Cor:AONT-Inverse}, where we swap the input and output columns to derive all $(1,3,3)$-AONTs.

We consider the array representation of a $(2,3,3)$-AONT, where each of the $3^3 = 27$ rows is of the form
\[
 ( (x_1, x_2 ,x_3), \phi(x_1, x_2 ,x_3)) = ((x_1, x_2 ,x_3),(x_{4}, x_{5} ,x_{6}))
\]
After fixing the ordering of the input columns in the array, there are initially $27!$ possible arrays to consider. However, most of them can be discarded as they fail to be unbiased with respect to the collections of columns specified in Theorem \ref{Thrm:array-equivalence}.

We use a branch-and-bound algorithm that explores possible row assignments in a depth-first manner and prunes the search tree using the constraints imposed by Corollary \ref{Cor:distance}. This approach reduces the effective search space from $27!$ to approximately $11!$. The reduction is achieved because the algorithm does not consider all choices for every row. As more rows are assigned, the constraints from Corollary \ref{Cor:distance} accumulate and heavily restrict the feasible output choices for the subsequent rows. Specifically, whenever the input vector of a row lies within Hamming distance of $s-t$ from many previously assigned input vectors, the corresponding output vector of the row must simultaneously avoid multiple Hamming spheres of radius $t$ centered at the previously assigned output vectors. The intersection of these constraints significantly reduces the number of branches to consider at each level.

The pseudocode in Algorithm \ref{algo1} describes an ordering on the input vectors that will allow early pruning of the search tree. Starting from an arbitrary input vector, the remaining vectors are ordered so that, at each level, the selected vector lies within Hamming distance of $s-t$ from as many previously placed vectors as possible.

Algorithm \ref{algo2} presents the general branch-and-bound procedure. At each level of the search, we consider all previously assigned input vectors whose Hamming distance from the current input vector is at most $s-t$, and determine the corresponding output vectors. We denote the set of such output vectors by $N$. By Corollary \ref{Cor:distance}, the infeasible output vectors at this level are simply those contained in the union of Hamming spheres of radius $t$ centered at the vectors in $N$. The Rust implementations of Algorithms \ref{algo1} and \ref{algo2} are provided here: \url{https://github.com/SPyCy-Lab/AONT-Search}.

\paragraph{Completeness of the enumeration.} The branch-and-bound procedure is exhaustive in the following sense: at every level, the only output assignments excluded are those provably infeasible by Corollary \ref{Cor:distance}, and every remaining assignment is explored recursively. The terminating condition ensures that only complete mappings are considered, and the recursion enumerates every such completion exactly once. Consequently, for any fixed ordering of the input column values, the algorithm yields every $(t,s,v)$-AONT extending the partial mapping consistent with that ordering.

\paragraph{Results.} Using this method, we enumerate a total of 192 linear $(2,3,3)$-AONTs and 4992 non-additive $(2,3,3)$-AONTs. An explicit example of a non-additive $(2,3,3)$-AONT arising from this enumeration is given in Table \ref{tab:non-linear-example-search}. 

\begin{proposition}\label{Prop:allt33attainable}
Every non-additive $(t,3,3)$-AONT is obtainable from some linear $(t,3,3)$-AONT
by a sequence of column-wise symbol permutations.
\end{proposition}
\begin{proof}
Verified by exhaustive enumeration: the 4992 non-linear $(2,3,3)$-AONTs
identified in Section \ref{sec:enum-all-general-t33} were checked computationally, and each
was found to arise from a linear $(2,3,3)$-AONT via column-wise symbol
permutations. The corresponding statement for $(1,3,3)$-AONTs follows
from Corollary \ref{Cor:AONT-Inverse}. The implementation of this search is provided here:

\url{https://github.com/SPyCy-Lab/AONT-Search}.\qed
\end{proof}

\paragraph{Scalability of the Branch-and-bound Enumeration.} While the branch-and-bound method substantially reduces the search space for small parameter sets, the reduction is not sufficient for larger cases. Larger values of $t$ improve the pruning efficiency of the search, since the Hamming spheres in Algorithm \ref{algo2} eliminate more candidate output vectors at each level of the search. To illustrate the scalability issue, we consider the next largest parameter set $(t,4,4)$ with the most favorable choice of $t$, namely $t=3$.

For the $(3,4,4)$-AONT case, there are $4^4=256$ input vectors. Under the pruning rule from Corollary \ref{Cor:distance}, a level with no previously placed neighbour may have up to $256$ feasible outputs. A level with one previously placed neighbour has roughly $81$ feasible outputs, with two previous neighbours roughly $16$, and with three or more previous neighbours often only one feasible output.

These estimates arise from excluding the Hamming sphere of radius $t$ around the already assigned neighbouring outputs. Since the Hamming sphere of radius $3$ in $\mathbb{F}_4^4$ contains $175$ vectors, a single neighbouring constraint removes $175$ candidate outputs from $256$ leaving only 81. These values are only approximate, since previously assigned output vectors must also be excluded. Nevertheless, they give a useful estimate of the effective branching factor. Using the ordering produced by Algorithm \ref{algo1}, we can count the number of previously placed neighbours at each level and multiply the corresponding estimated branching factors. For the case of $(3,4,4)$, this gives an effective search complexity on the order of $22!$ to $23!$, which remains computationally infeasible. Thus, while the branch-and-bound algorithm avoids the full $256!$ search, it does not scale sufficiently for larger parameter sets.

\begin{algorithm}
\caption{Ordering of Input Vectors by Previously Placed Neighbours}
\label{algo1}
\begin{algorithmic}[1]
\State \textbf{Input:} 
\State\textit{$T$ : the set of all the vectors in $\mathbb{F}_q^s$}
\State\textit{$H$: a mapping, where $H(x)=S_{s-t}(x)\setminus\{x\}$}
\State \textbf{Output:} 
\State\textit{An ordered list $L=[(x_i,N_i)]$, where
$N_i = H(x_i)\cap \{x_1,\ldots,x_{i-1}\}$}

\State Let $x_1$ be the all-zero vector
\State $L \gets [(x_1,\emptyset)]$
\State $P \gets \{x_1\}$

\While{$P \neq T$}
    \State Choose $\quad  x \in T\setminus P
        \quad \text{maximizing} \quad
        |H(x)\cap P|$
    \State $N \gets H(x)\cap P$
    \State Append $(x,N)$ to $L$
    \State $P \gets P\cup\{x\}$
\EndWhile

\State \Return $L$
\end{algorithmic}
\end{algorithm}

\begin{algorithm}
\caption{Branch-and-Bound Search for General $(t,s,q)$-AONTs}
\label{algo2}
\begin{algorithmic}[1]
\State\textbf{Input:}
\State\textit{$(t,s,q)$ : the AONT parameters}
\State\textit{$[(x_i,N_i)]_{i=1}^{q^s}$ : ordered inputs from Algorithm 1, where
$N_i$ is the set of previously placed neighbours of $x_i$}
\State\textit{$U$ : the set of all the vectors in $\mathbb{F}_q^s$}
\State\textit{$\phi$ : the current partial mapping}
\State \textbf{Output:} All general $(t,s,q)$-AONTs extending $\phi$.

\If{$\phi$ is defined on all vectors in $U$}
    \State Output $\phi$
    \State \Return
\EndIf

\State Let $(x_i,N_i)$ be the next pair in the given ordering.

\State $F$ $\gets$ outputs already used by $\phi$

\For{each neighbour $x_j \in N_i$}
    \State Add all vectors in $S_{t}(\phi(x_j))$ to $F$
\EndFor

\For{each $y \in U \setminus$ $F$}
    \State Recursively search using the extended mapping
    \[
        \phi' = \phi \cup \{x_i \mapsto y\}.
    \]
\EndFor

\end{algorithmic}
\end{algorithm}

\begin{table}
\centering
\caption{A non-additive $(2,3,3)$-AONT over $\mathbb{F}_3$}
\label{tab:non-linear-example-search}
\fontsize{10}{11}\selectfont
\renewcommand{\arraystretch}{1.15}
\begin{tabular}{|c|c c c|c c c||c|c c c|c c c|}
\hline
No.& $x_1$ & $x_2$ & $x_3$ & $x_4$ & $x_5$ & $x_6$ & No.& $x_1$ & $x_2$ & $x_3$ & $x_4$ & $x_5$ & $x_6$ \\
\hline
1 & $0$ & $0$ & $0$ & $0$ & $0$ & $1$ &  
2 & $0$ & $0$ & 2 & $1$ & $1$ & $0$ \\  
\hline
3 & $0$ & $0$ & $1$ & $2$ & $2$ & $2$ &  
4 & $2$ & $0$ & $2$ & $0$ & $0$ & $2$ \\  
\hline
5 & $2$ & $0$ & $1$ & $1$ & $1$ & $1$ &  
6 & $2$ & $0$ & $0$ & $2$ & $2$ & $0$ \\  
\hline
7 & $1$ & $0$ & $2$ & $2$ & $2$ & $1$ &  
8 & $1$ & $0$ & $1$ & $0$ & $0$ & $0$ \\  
\hline
9 & $1$ & $0$ & $0$ & $1$ & $1$ & $2$ &  
10 & $2$ & $2$ & $2$ & $1$ & $2$ & $0$ \\  
\hline
11 & $2$ & $2$ & $1$ & $2$ & $0$ & $2$ &  
12 & $2$ & $2$ & $0$ & $0$ & $1$ & $1$ \\  
\hline
13 & $2$ & $1$ & $2$ & $2$ & $1$ & $1$ &  
14 & $2$ & $1$ & $0$ & $1$ & $0$ & $2$ \\  
\hline
15 & $2$ & $1$ & $1$ & $0$ & $2$ & $0$ &  
16 & $1$ & $2$ & $2$ & $0$ & $1$ & $2$ \\  
\hline
17 & $1$ & $2$ & $1$ & $1$ & $2$ & $1$ &  
18 & $1$ & $2$ & $0$ & $2$ & $0$ & $0$ \\  
\hline
19 & $1$ & $1$ & $2$ & $1$ & $0$ & $0$ &  
20 & $1$ & $1$ & $1$ & $2$ & $1$ & $2$ \\  
\hline
21 & $1$ & $1$ & $0$ & $0$ & $2$ & $1$ &  
22 & $0$ & $2$ & $0$ & $1$ & $2$ & $2$ \\  
\hline
23 & $0$ & $2$ & $2$ & $2$ & $0$ & $1$ &  
24 & $0$ & $2$ & $1$ & $0$ & $1$ & $0$ \\  
\hline
25 & $0$ & $1$ & $1$ & $1$ & $0$ & $1$ &  
26 & $0$ & $1$ & $0$ & $2$ & $1$ & $0$ \\  
\hline
27 & $0$ & $1$ & $2$ & $0$ & $2$ & $2$ & & & & & &  &\\  
\hline

\end{tabular}

\vspace{0.3cm}
\textbf{Evidence of non-additivity.}\\
Let $x=(0,0,0)$ and $y=(2,2,2)$. Then
$\phi(x)+\phi(y)=(1,2,1)$, whereas $\phi(x+y)=(1, 2, 0)$, showing that $\phi(x+y)\neq \phi(x)+\phi(y)$ and hence the mapping is non-additive.
\end{table}

\section{Constructing non-additive and non-affine AONTs }\label{sec:const-non-linear}

We now present a method to construct a non-additive $(t,s,q)$-AONT, from a linear one. We start with the unbiased array representation of a linear AONT and apply a symbol permutation using Corollary \ref{Cor:sym_permutation} to construct a non-additive AONT.

\begin{construction} \label{Con:nonlinear-construction}
    Let $\phi : \mathcal{X}^s \to \mathcal{X}^s$ be a linear $(t,s,q)$-AONT, where $q$ is a prime power, and let $A$ denote its array representation, where each row of $A$ is of the form 
    \[
     ( (x_1, ... ,x_s), \phi(x_1, ... ,x_s)) = (x_1, ... ,x_s,x_{s+1}, ... ,x_{2s})
    \]

    We consider the cases of even $q$ and odd $q$ separately.
    \begin{case}[Odd $q$]
        Choose two rows, $\mathbf{a}$ and $\mathbf{b}$, from $A$ such that their output entries differ in at least one coordinate, and in at least one such coordinate both entries are nonzero. Without loss of generality, assume that this occurs in the last output column.

        We now have two rows as such 
    \begin{align*}
     \mathbf{a}&=(a_1, ... ,a_{2s-1} ,\alpha), \text{ and}\\
     \mathbf{b}&=(b_1, ... ,b_{2s-1} ,\beta)
     \end{align*}
    where $\mathbf{a}$ and $\mathbf{b}$ are arbitrary input vectors. By linearity, it follows that
    \[
     \mathbf{c} = (a_1+b_1, ... ,a_{2s-1}+b_{2s-1}, \gamma= \alpha +\beta)
    \]

    Note that operations are done over $\mathbb{F}_q$, and that $\alpha = \gamma -\beta$.
    
    Now apply a symbol swap permutation to the last column of $A$ swapping all occurrences of  $\alpha$ and $\gamma$, to obtain the new array $A'$. 

    The array $A'$ represents a $(t,s,q)$-AONT $\phi'\neq \phi$. 
    \end{case}

    \begin{case}[Even $q$]
    Suppose $\theta$\footnote{The more common symbols for the root of the irreducible polynomial $x$ and $\alpha$ are already used elsewhere for other purposes.} is the root of the irreducible polynomial used to generate $\mathbb{F}_{2^p}, p>1$. Choose two rows, $\mathbf{a}$ and $\mathbf{b}$, from $A$ such that there exists a coordinate $i$ in their output entries where $\mathbf{a}_i = 0$ and $\mathbf{b}_i = 1$. Without loss of generality, assume that this occurs in the last output column.

Note that operations are done over $\mathbb{F}_{2^p}$.
\begin{align*}
    \mathbf{a}&= (a_1, \dots , a_{2s-1}, 0), \text{ and} \\
    \mathbf{b}&= (b_1, \dots , b_{2s-1}, 1)
\end{align*}

 where $\mathbf{a}$ and $\mathbf{b}$ are arbitrary input vectors. By linearity, it follows that
    \[
     \mathbf{c} = (a_1+b_1, ... ,a_{2s-1}+b_{2s-1}, 1).
    \]

Consider the symbol permutation $\pi$ defined as below.
\[
\pi(0) = \theta, \pi(1) = 0, \pi(\theta) = 1, \text{ and } \pi(x) = x, \forall x\in \mathbb{F}_{2^p}\setminus\{0,1,\theta\}.
\]
 Now we apply $\pi$ to the last column of $A$, to obtain the new array $A'$. 
    \end{case}

Again in this case, the array $A'$ represents a $(t,s,q)$-AONT $\phi'\neq \phi$.

\end{construction}

\begin{theorem}\label{Thrm:construction-correctness}
The output of Construction \ref{Con:nonlinear-construction} is a non-additive AONT.
\end{theorem}

 \begin{proof}
 In both cases, by Corollary \ref{Cor:sym_permutation}, the array $A'$ represents a $(t,s,q)$-AONT $\phi'$. 
 
 In the odd $q$ case, $\phi'$ includes the rows
\begin{align*}
    \mathbf{a'}&=(a_1, ... ,a_{2s-1},\gamma),\\
     \mathbf{b'}&=(b_1, ... ,b_{2s-1},\beta), \\
     \mathbf{c'}&= (a_1+b_1, ... ,a_{2s-1}+b_{2s-1} ,\alpha)
\end{align*}
     
The last coordinate of $\phi'(\mathbf{a}_{[1,s]}) + \phi'(\mathbf{b}_{[1,s]})$ is $\gamma + \beta = \alpha +2\beta$, while the last coordinate of $\phi'(\mathbf{a}_{[1,s]} + \mathbf{b}_{[1,s]})$ is $\alpha$. Since $\beta\neq 0$ (recall that in Case 1 we chose both $\alpha$ and $\beta$ to be nonzero) and $q$ is odd, these two coordinates differ, so 
    \[
     \phi'(\mathbf{a}_{[1,s]}) + \phi'(\mathbf{b}_{[1,s]}) \neq\phi'(\mathbf{a}_{[1,s]} + \mathbf{b}_{[1,s]}). 
    \]
    
 In the even $q$ case, $\phi'$ includes the rows
 \begin{align*}
     \mathbf{a'}&= (a_1, \dots , a_{2s-1}, \theta)\\
     \mathbf{b'}&= (b_1, \dots , b_{2s-1}, 0) \\
     \mathbf{c'}&= (a_1+b_1, ... ,a_{2s-1}+b_{2s-1} , 0)
 \end{align*}

The last coordinate of $\phi'(\mathbf{a}_{[1,s]}) + \phi'(\mathbf{b}_{[1,s]})$ is $\theta + 0 = \theta$, while the last coordinate of $\phi'(\mathbf{a}_{[1,s]} + \mathbf{b}_{[1,s]})$ is $0$. Since $\theta\neq 0$, these two coordinates differ, so
    \[
     \phi'(\mathbf{a}_{[1,s]}) + \phi'(\mathbf{b}_{[1,s]}) \neq\phi'(\mathbf{a}_{[1,s]} + \mathbf{b}_{[1,s]}). 
    \]
    In both cases, the resulting mapping $\phi'$ is non-additive. \qed
 \end{proof}

It is worth noting that $(t,s,2)$-AONTs do not exist, and the construction above can be applied to all existing linear AONTs.
\begin{remark}
Multiple symbol permutations can be applied to a single column, as long as each new permutation acts on a set of symbols disjoint from those involved in any previous permutation. In particular, if $\alpha$ and $\gamma$ were swapped, we require that subsequent permutations do not modify $\alpha$, $\beta$, and $\gamma = \alpha +\beta$. In addition, symbol permutations can be applied independently to all  columns. This enables the construction of additional non-additive variants of the mapping. We study this in Section \ref{sec:measuring-non-linearity}.
\end{remark}

We now show that the same construction yields non-affine AONTs, when the underlying field is large enough. 

\begin{theorem}\label{Thrm:construction-is-non-affine}
The output of Construction \ref{Con:nonlinear-construction} is a non-affine AONT for every prime power $q \geq 5$.
\end{theorem}

\begin{proof}
Let $\phi$ be the original linear AONT, and let $\phi'$ be the AONT obtained from Construction \ref{Con:nonlinear-construction}. Suppose for the sake of contradiction, that $\phi'$ is affine. Then there exists a fixed vector $\mathbf{k} \in \mathbb F_q^s$ such that
\[
\phi'(\mathbf{x}) + \phi'(\mathbf{y}) - \phi'(\mathbf{x+y}) = \mathbf{k}
\]
for every $\mathbf{x},\mathbf{y}\in\mathbb F_q^s$.
 \begin{case}[Odd q]
 Consider the rows $\mathbf{a}$, $\mathbf{b}$, and $\mathbf{c}$ from the construction. Choose an additional row
 \[
 \mathbf{d}=(d_1, ... ,d_{2s-1},\delta),
 \]
 where $\delta \notin \{0,\alpha,\beta,\gamma\}$. Such a symbol exists since $q \geq 5$. By linearity the array $A$ also contains the row
 \[
 \mathbf{e}=(a_1+d_1, ... ,a_{2s-1} +d_{2s-1},\alpha+\delta),
 \]
 Since neither $\delta$ nor $\alpha+\delta$ belongs to $\{\alpha,\gamma\}$, the rows $\mathbf{e}$ and $\mathbf{d}$ remain unchanged after swapping $\alpha$ and $\gamma$.

 Considering the rows $\mathbf{a'}$, $\mathbf{b'}$, and $\mathbf{c'}$, the last coordinate of $\mathbf{k}$ must satisfy
 \[
 k_s=\gamma+\beta-\alpha=2\beta.
 \]
 Similarly, considering  the rows $\mathbf{a'}$, $\mathbf{d}$, and $\mathbf{e}$, we obtain
 \[
  k_s=\gamma+\delta-(\alpha+\delta)=\gamma - \alpha = \beta.
 \]
 Since $\beta\neq 0$, we have $2\beta\neq\beta$, contradicting the assumption that $\mathbf{k}$ is fixed.
 \end{case}

 \begin{case}[Even q]
 Consider the rows $\mathbf{a}$, $\mathbf{b}$, and $\mathbf{c}$ from the construction. Choose an additional row
 \[
 \mathbf{d}=(d_1,\ldots,d_{2s-1},\delta),
 \]
 where $\delta\notin\{0,1,\theta,1+\theta\}$. Such a symbol exists since $q\geq 8$. By linearity, the array $A$ also contains the row
 \[
 \mathbf{e}=(b_1+d_1,\ldots,b_{2s-1}+d_{2s-1},1+\delta).
 \]
 Since neither $\delta$ nor $1+\delta$ belongs to $\{0,1,\theta\}$, the rows $\mathbf{d}$ and $\mathbf{e}$ remain unchanged after applying the symbol permutation.
 
 Considering the rows $\mathbf{a'}$, $\mathbf{b'}$, and $\mathbf{c'}$, the last coordinate of $\mathbf{k}$ must satisfy
 \[
 k_s=\theta+0-0=\theta.
 \]
 Similarly, considering the rows $\mathbf{b'}$, $\mathbf{d}$, and $\mathbf{e}$, we obtain
 \[
 k_s=0+\delta-(1+\delta)= -1 = 1.
 \]
 Since $\theta \neq 1$, this contradicts the assumption that $\mathbf{k}$ is fixed.
 \end{case}
Therefore, in both cases, no such fixed vector $\mathbf{k}$ exists, and hence $\phi'$ is non-affine.
 \qed
\end{proof}

\subsection{Example construction of a non-additive AONT}\label{subsec4}

Select two rows from a linear $(2,3,3)$-AONT over the alphabet $\{0,1,2\}$ such that their last output entries are differing and nonzero. By applying a symbol permutation to the last column of the array that swaps the symbols \textit{0} and \textit{2}, we obtain the non-additive $(2,3,3)$-AONT shown in Table \ref{tab:non-linear-example-construction}.

\begin{figure}[htbp]
\centering
\caption{Selecting rows from a linear $(2,3,3)$-AONT}
\begin{tikzpicture}[>=Stealth]
\node[anchor=north west] (full) at (0,0) {%

\begin{tabular}{|c|c c c|c c c||c|c c c|c c c|}
\hline
No. & $x_1$ & $x_2$ & $x_3$ & $x_4$ & $x_5$ & $x_6$ & No. & $x_1$ & $x_2$ & $x_3$ & $x_4$ & $x_5$ & $x_6$ \\
\hline
1 & $0$ & $0$ & $0$ & $0$ & $0$ & $0$ &  
2 & $0$ & $0$ & $1$ & $2$ & $2$ & $2$ \\  
\hline
3 & $0$ & $0$ & $2$ & $1$ & $1$ & $1$ &  
4 & $0$ & $1$ & $0$ & $2$ & $1$ & $2$ \\  
\hline
5 & $0$ & $1$ & $1$ & $1$ & $0$ & $1$ &  
6 & $0$ & $1$ & $2$ & $0$ & $2$ & $0$ \\  
\hline
7 & $0$ & $2$ & $0$ & $1$ & $2$ & $1$ &  
8 & $0$ & $2$ & $1$ & $0$ & $1$ & $0$ \\  
\hline
9 & $0$ & $2$ & $2$ & $2$ & $0$ & $2$ &  
10 & $1$ & $0$ & $0$ & $1$ & $1$ & $2$ \\  
\hline
11 & $1$ & $0$ & $1$ & $0$ & $0$ & $1$ &  
12 & $1$ & $0$ & $2$ & $2$ & $2$ & $0$ \\  
\hline
13 & $1$ & $1$ & $0$ & $0$ & $2$ & $1$ &  
14 & $1$ & $1$ & $1$ & $2$ & $1$ & $0$ \\  
\hline
15 & $1$ & $1$ & $2$ & $1$ & $0$ & $2$ &  
16 & $1$ & $2$ & $0$ & $2$ & $0$ & $0$ \\  
\hline
17 & $1$ & $2$ & $1$ & $1$ & $2$ & $2$ &  
18 & $1$ & $2$ & $2$ & $0$ & $1$ & $1$ \\  
\hline
19 & $2$ & $0$ & $0$ & $2$ & $2$ & $1$ &  
20 & $2$ & $0$ & $1$ & $1$ & $1$ & $0$ \\  
\hline
21 & $2$ & $0$ & $2$ & $0$ & $0$ & $2$ &  
22 & $2$ & $1$ & $0$ & $1$ & $0$ & $0$ \\  
\hline
23 & $2$ & $1$ & $1$ & $0$ & $2$ & $2$ &  
24 & $2$ & $1$ & $2$ & $2$ & $1$ & $1$ \\  
\hline
25 & $2$ & $2$ & $0$ & $0$ & $1$ & $2$ &  
26 & $2$ & $2$ & $1$ & $2$ & $0$ & $1$ \\  
\hline
27 & $2$ & $2$ & $2$ & $1$ & $2$ & $0$ & & & & & & &  \\ 
\hline
\end{tabular}
};

\node[anchor=north west, xshift=1cm] (sel) at (full.north east) {%
\begin{tabular}{|c|c|c c c|c c c|}
\hline
No.&$Vec$ & $x_1$ & $x_2$ & $x_3$ & $x_4$ & $x_5$ & $x_6$ \\
\hline
$21$&$\mathbf{a}$ & $2$& $0$ & $2$ & $0$ & $0$ & $2 = \alpha $ \\
$19$&$\mathbf{b}$ & $2$& $0$& $0$ & $2$ & $2$ & $1 = \beta $ \\
$12 $&$\mathbf{a}+ \mathbf{b}$ & $1$& $0$& $2$& $2$& $2$& $0 = \gamma $ \\
\hline
\end{tabular}
};

\end{tikzpicture}
\end{figure}

\begin{table}[!htbp]
\centering
\caption{$\phi'$ a non-additive $(2,3,3)$-AONT resulting from the symbol permutation}
\label{tab:non-linear-example-construction}
\fontsize{10}{11}\selectfont
\renewcommand{\arraystretch}{1.15}
\begin{tabular}{|c|c c c|c c c||c|c c c|c c c|}
\hline
No. & $x_1$ & $x_2$ & $x_3$ & $x_4$ & $x_5$ & $x_6$ & No. & $x_1$ & $x_2$ & $x_3$ & $x_4$ & $x_5$ & $x_6$ \\
\hline
1 & $0$ & $0$ & $0$ & $0$ & $0$ & $2$ &  
2 & $0$ & $0$ & $1$ & $2$ & $2$ & $0$ \\  
\hline
3 & $0$ & $0$ & $2$ & $1$ & $1$ & $1$ &  
4 & $0$ & $1$ & $0$ & $2$ & $1$ & $0$ \\  
\hline
5 & $0$ & $1$ & $1$ & $1$ & $0$ & $1$ &  
6 & $0$ & $1$ & $2$ & $0$ & $2$ & $2$ \\  
\hline
7 & $0$ & $2$ & $0$ & $1$ & $2$ & $1$ &  
8 & $0$ & $2$ & $1$ & $0$ & $1$ & $2$ \\  
\hline
9 & $0$ & $2$ & $2$ & $2$ & $0$ & $0$ &  
10 & $1$ & $0$ & $0$ & $1$ & $1$ & $0$ \\  
\hline
11 & $1$ & $0$ & $1$ & $0$ & $0$ & $1$ &  
12 & $1$ & $0$ & $2$ & $2$ & $2$ & $2$ \\  
\hline
13 & $1$ & $1$ & $0$ & $0$ & $2$ & $1$ &  
14 & $1$ & $1$ & $1$ & $2$ & $1$ & $2$ \\  
\hline
15 & $1$ & $1$ & $2$ & $1$ & $0$ & $0$ &  
16 & $1$ & $2$ & $0$ & $2$ & $0$ & $2$ \\  
\hline
17 & $1$ & $2$ & $1$ & $1$ & $2$ & $0$ &  
18 & $1$ & $2$ & $2$ & $0$ & $1$ & $1$ \\  
\hline
19 & $2$ & $0$ & $0$ & $2$ & $2$ & $1$ &  
20 & $2$ & $0$ & $1$ & $1$ & $1$ & $2$ \\  
\hline
21 & $2$ & $0$ & $2$ & $0$ & $0$ & $0$ &  
22 & $2$ & $1$ & $0$ & $1$ & $0$ & $2$ \\  
\hline
23 & $2$ & $1$ & $1$ & $0$ & $2$ & $0$ &  
24 & $2$ & $1$ & $2$ & $2$ & $1$ & $1$ \\  
\hline
25 & $2$ & $2$ & $0$ & $0$ & $1$ & $0$ &  
26 & $2$ & $2$ & $1$ & $2$ & $0$ & $1$ \\  
\hline
27 & $2$ & $2$ & $2$ & $1$ & $2$ & $2$ & & & & & & &  \\ 
\hline

\end{tabular}
\vspace{0.3cm}

\parbox{0.95\linewidth}{
\centering
\textbf{Evidence of non-additivity.}\\
We consider the previously selected rows, $\mathbf{a}=(2,0,2)$ and $\mathbf{b}=(2,0,0)$. Then
$\phi'(\mathbf{a})+\phi'(\mathbf{b})=(2,2,1)$, whereas $\phi'(\mathbf{a} + \mathbf{b})=(2,2,2)$, showing that $\phi'(\mathbf{a} + \mathbf{b})\neq \phi'(\mathbf{a})+\phi'(\mathbf{b})$ and hence the mapping is non-additive.
}
\end{table}

\section{Measuring Non-linearity}\label{sec:measuring-non-linearity}

In this section, we analyze the non-linearity of our proposed construction method and the enumerated $(t,3,3)$-AONTs, using analytical and computational methods. We evaluate the differential $\Delta_F$-uniformity and Walsh spectra exhaustively over the vector space $\mathbb{F}_q^s$ for some non-affine AONTs, and investigate the effect of column-wise symbol permutations on these measures. 
 
 As presented in Proposition \ref{Prop:allt33attainable} all the non-additive $(t,3,3)$-AONTs are attainable from the linear ones. Furthermore, all such $(t,3,3)$-AONTs were found to be affine, attaining maximal differential $\Delta_F$-uniformity and maximal Walsh magnitude. Despite being non-additive, these mappings did not exhibit cryptographic non-linearity. 

While measuring different constructions the computations revealed which type of column-permutations reduce differential $\Delta_F$-uniformity and Walsh magnitude. We begin by introducing the notion of difference-splitting permutations and examining how applying them to the output coordinates of a linear AONT affects its differential uniformity and $W_{max}$. We first show how these permutations split the differential counts associated with a linear mapping and identify cases in which these counts remain unchanged. Building on these observations, we then establish that applying difference-splitting permutations to all output coordinates reduces the overall differential uniformity, although this reduction is limited by a lower bound of $q^{s-1}$. Similarly we also show that applying difference-splitting permutations to all output coordinates reduces $W_{max}$. Finally, we turn our attention to permutations applied on the input side and show that their effect on differential uniformity and $W_{max}$ can be characterized by the same framework developed for output side permutations.

We begin with the definition of a Difference-splitting permutation.
\begin{definition}\label{def:diff-split}
Let $\pi:\mathbb{F}_q\to\mathbb{F}_q$ be a permutation, where $q=p^m$. We say that $\pi$ is
difference-splitting if both of the following 
conditions hold:
\begin{enumerate}
    \item For every nonzero $d\in\mathbb{F}_q$,
    the set
    \[
    \{\pi(x+d)-\pi(x):x\in\mathbb{F}_q\}
    \]
    contains more than one value.

    \item For every $a\in\mathbb{F}_q$ and
    nonzero $b\in\mathbb{F}_q$, the set
    \[
    \{\operatorname{Tr}(b\pi(x)-ax):
    x\in\mathbb{F}_q\}
    \]
    contains more than one value, where
    $\operatorname{Tr}:\mathbb{F}_q\to
    \mathbb{F}_p$ is the absolute trace.
\end{enumerate}
\end{definition}

\begin{remark}
    Suppose $p$ is prime. A permutation $\pi: \mathbb{F}_p \to \mathbb{F}_p$ is difference-splitting if and only if $\pi$ is non-affine over $\mathbb{F}_p$, i.e., $\pi(x)\neq ax+b$ for any choice of $a, b \in \mathbb{F}_p$. 
    
    Indeed, if $\pi(x) = ax+b$ then $\pi(x+d)-\pi(x) = ad$ is constant in $x$ for every $d$. Conversely, if $\pi(x+d)-\pi(x)$ is constant in $x$ for every $d$, then $\pi(x)-\pi(0)$ is additive, and since every additive map $\mathbb{F}_p\to\mathbb{F}_p$ is multiplication by a scalar, $\pi$ is affine. Thus non-affinity implies the first condition, and also the second, since over prime fields every nonzero trace component of a non-affine permutation is non-affine.

    The converse does not hold over extension fields $\mathbb{F}_{p^m}$ with $m>1$, since additive permutations need not be of the form $ax+b$.
\end{remark}

We now show that difference-splitting permutations split the differential counts associated with a linear AONT.

\begin{proposition}\label{Prop:diff_split-reduces}
Let
\[
\phi:\mathbb{F}_q^s\to\mathbb{F}_q^s
\]
be a linear AONT, and let $\phi'$ be obtained by applying a
difference-splitting permutation $\pi$ to a single output coordinate,
without loss of generality let it be the last coordinate. 
Suppose
\[
a\in\mathbb{F}_q^s
\]
is nonzero and the last coordinate of $\phi(a)$ is nonzero. Then the
$q^s$ solutions contributing to
\[
\delta_\phi(a,\phi(a))
\]
are partitioned among multiple values of $b$ in the differential
counts of $\phi'$.
\end{proposition}

\begin{proof}
Since $\phi$ is linear,
\[
\phi(x+a)-\phi(x)=\phi(a)
\]
for all
\[
x\in\mathbb{F}_q^s.
\]
Thus, before the permutation, every $x\in\mathbb{F}_q^s$ contributes
to the same differential count
\[
\delta_\phi(a,\phi(a)),
\]
and therefore
\[
\delta_\phi(a,\phi(a))=q^s.
\]

Let $d\neq0$ denote the last coordinate of $\phi(a)$. Then for every
$x$,
\[
\phi(x+a)_s-\phi(x)_s=d.
\]

After applying the permutation $\pi$ to the last coordinate, the new
difference in that coordinate becomes
\[
\pi(\phi(x+a)_s)-\pi(\phi(x)_s).
\]

Since $\pi$ is difference-splitting, the set
\[
\{\pi(y+d)-\pi(y):y\in\mathbb{F}_q\}
\]
contains more than one value. Hence the original constant difference
$d$ is transformed into multiple distinct differences as $x$ varies.
All other coordinates remain unchanged, so the values that originally
contributed to $\delta_\phi(a,\phi(a))$ now contribute to several distinct
counts $\delta_{\phi'}(a,\phi(a))$, where the corresponding $b$'s differ only
in the last coordinate. In particular, $\delta_{\phi'}(a,\phi(a))< q^s$. \qed
\end{proof}

Difference-splitting permutations applied on a single-column only reduce the differential count for $\delta(a,b)$ when $\phi(a)$ is nonzero in that column. The following proposition explains the case in which single-column permutations cannot reduce the $\Delta_F$-uniformity count for AONTs.

\begin{proposition}\label{Prop:single_col_limit}
Let 
\[
\phi : \mathbb{F}_q^s \to \mathbb{F}_q^s
\]
be a linear AONT mapping, and let 
\[
\phi' = \pi \circ \phi
\]
where $\pi$ represents a symbol permutation applied to a single column without loss of generality, we assume it to be the last output column.

Consider a nonzero
\[
a \in \mathbb{F}_q^s 
\]
such that the last coordinate of $\phi(a)$ is zero. Then
\[
\delta_{\phi'}(a,\phi(a)) = \delta_{\phi}(a,\phi(a))
\]
In particular, the corresponding differential count remains unchanged for the pair $(a,\phi(a))$ under the permutation.
\end{proposition}

\begin{proof}
Since $\phi$ is linear,
\[
\phi(x+a) - \phi(x) = \phi(a)
\]
for all 
\[
x \in \mathbb{F}_q^s 
\]
As the last coordinate of $\phi(a)$ is zero, the last coordinates of $\phi(x+a)$ and $\phi(x)$ are equal for every $x$.

Applying the symbol permutation to the last coordinate therefore maps both terms to the same symbol, leaving the last-coordinate difference unchanged. Hence,
\[
\phi'(x+a) - \phi'(x) = \phi(a)
\]
holds for the same set of $x$ as before the permutation, showing
\[
\delta_{\phi'}(a,\phi(a)) = \delta_{\phi}(a,\phi(a)).
\] 
\qed
\end{proof}

Considering the preceding two results, we present a proposition characterizing how $\Delta_F$-uniformity is affected when difference-splitting permutations are applied on less than $s$ columns.

\begin{proposition}\label{prop:apply-on-all-columns}
Suppose
\[
\phi:\mathbb{F}_q^s\to\mathbb{F}_q^s
\]
is a linear AONT, and $\phi'$ is obtained by applying difference-splitting permutations to fewer than $s$ output coordinates of $\phi$. Then 
\[
\Delta_{\phi'} = q^s.
\]
\end{proposition}
\begin{proof}
Let $i$ be an output coordinate on which no difference-splitting permutation is applied. Since $\phi$ is a bijection over $\mathbb{F}_q^s$, 
we can consider a nonzero
\[
a \in \mathbb{F}_q^s
\]
such that only the $i$-th coordinate of $\phi(a)$ is nonzero. i.e.
\[
\phi(a)_{i} \neq 0
\]
and 
\[
\phi(a)_{j} = 0, \quad  j \neq i
\]
Since all other coordinates of $\phi(a)$ are zero, we know from Proposition \ref{Prop:single_col_limit} that any difference-splitting permutations applied to those coordinates will not affect $\delta_{\phi'}(a,\phi(a))$ thus
\[
\delta_{\phi'}(a,\phi(a)) = \delta_{\phi}(a,\phi(a)) = q^s
\]
Therefore,
\[
\Delta_{\phi'} = q^s.
\]
\qed
\end{proof}

We next show that applying difference-splitting permutations to all $s$ output columns will reduce $\delta(a,b)$ for all $(a,b)$, resulting in reduced  $\Delta_F$-uniformity for the entire AONT.

\begin{proposition}\label{prop:all-column-permutation}
Let
\[
\phi:\mathbb{F}_q^s\to\mathbb{F}_q^s
\]
be a linear AONT, and let $\phi'$ be obtained by applying
difference-splitting permutations independently to all output
coordinates. Then for every nonzero
\[
a\in\mathbb{F}_q^s,
\]
the differential count corresponding to
\[
(a,\phi(a))
\]
splits among multiple values of $b$ under $\phi'$.
\end{proposition}

\begin{proof}
Since $\phi$ is linear, if $a\neq0$ then $\phi(a)\neq0$.

Hence, for every nonzero $a$, at least one coordinate of $\phi(a)$ is
nonzero.

Fix such a coordinate $j$, and let
\[
d=\phi(a)_j\neq0.
\]
Since the permutation applied to coordinate $j$ is
difference-splitting, the set
\[
\{\pi_j(y+d)-\pi_j(y):y\in\mathbb{F}_q\}
\]
contains more than one value. Therefore the constant difference in
coordinate $j$ is partitioned into multiple distinct differences after
the permutation.

As in the previous proposition \ref{Prop:diff_split-reduces}, this causes the original differential count corresponding to
\[
(a,\phi(a))
\]
to split into multiple values of $b$. Since every nonzero $a$ has at least
one nonzero coordinate in $\phi(a)$, this splitting occurs for every
nonzero $a$, and therefore $\Delta_{\phi'} < q^s$. \qed
\end{proof}

We next establish a lower bound on the differential uniformity when permutations are applied only to the output coordinates of a linear AONT.

\begin{proposition}\label{prop:output-lower-bound}
Suppose
\[
\phi:\mathbb{F}_q^s\to\mathbb{F}_q^s
\]
is a linear AONT, and $\phi'$ is obtained by applying symbol permutations to all $s$ output coordinates of $\phi$. Then
\[
\Delta_{\phi'}\ge q^{s-1}.
\]
\end{proposition}

\begin{proof}
Since $\phi$ is a bijection over $\mathbb{F}_q^s$, we can consider a nonzero
\[
a\in\mathbb{F}_q^s
\]
such that only the $i$-th coordinate of $\phi(a)$ is nonzero, i.e.,
\[
\phi(a)_i\neq 0
\]
and
\[
\phi(a)_j=0,\quad j\neq i.
\]

Since $\phi$ is linear,
\[
\phi(x+a)-\phi(x)=\phi(a)
\]
for every $x\in\mathbb{F}_q^s$.

After applying the symbol permutations, the output difference can vary only in the $i$-th coordinate, since all other coordinates have zero difference.

Therefore, for this fixed $a$, there are at most $q$ distinct output differences $b$.

Since there are $q^s$ possible input vectors $x$, by the pigeonhole principle, at least one output difference must occur at least $q^{s-1}$ times. Thus,
\[
\Delta_{\phi'}\ge q^{s-1}.
\]
\qed
\end{proof}

We now investigate how applying difference-splitting
permutations to the output coordinates of a linear
AONT affects its Walsh transform. In particular, we
first show that such permutations reduce the magnitude
of Walsh coefficients associated with the permuted
coordinates. 

\begin{proposition}\label{Prop:diff_split-reduces-walsh}
Let $\phi:\mathbb{F}_q^s\to\mathbb{F}_q^s$ be a
linear AONT, and let $\phi'$ be obtained by applying
a difference-splitting permutation $\pi$ to its
$i$-th output coordinate. Then, for every
$\mathbf{a}\in\mathbb{F}_q^s$ and every nonzero
$\mathbf{b}\in\mathbb{F}_q^s$ supported only on
coordinate $i$,
\[
|W_{\phi'}(\mathbf{a},\mathbf{b})|<q^s.
\]
\end{proposition}
\begin{proof}
Let $\mathbf{b}$ be supported only on coordinate
$i$, with $b_i\neq 0$. Then
\[
W_{\phi'}(\mathbf{a},\mathbf{b})
=
\sum_{\mathbf{x}\in\mathbb{F}_q^s}
\omega^{\operatorname{Tr}
(b_i\pi(\phi(\mathbf{x})_i)
-\langle\mathbf{a},\mathbf{x}\rangle)}
\]

Now suppose, for a contradiction, that
$|W_{\phi'}(\mathbf{a},\mathbf{b})|=q^s$
for some $\mathbf{a} \in \mathbb F_q^s$.

Since $|W_{\phi'}(\mathbf{a},\mathbf{b})|=q^s$, there exists a constant $c\in\mathbb{F}_p$ such that
\[
\operatorname{Tr}
\bigl(b_i\pi(\phi(\mathbf{x})_i)
-\langle\mathbf{a},\mathbf{x}\rangle\bigr)
=c
\]
for every $\mathbf{x}\in\mathbb{F}_q^s$.

Since $\phi$ is invertible and linear, there
exists $\mathbf{v}\in\mathbb{F}_q^s$ such that
$\phi(\mathbf{v})=\mathbf{e}_i$.

For any $z\in\mathbb{F}_q$, linearity gives
$\phi(z\mathbf{v})=z\mathbf{e}_i$.
Substituting $\mathbf{x}=z\mathbf{v}$ yields
\[
\operatorname{Tr}
\bigl(b_i\pi(z)-z\langle\mathbf{a},\mathbf{v}\rangle
\bigr)=c
\]
for every $z\in\mathbb{F}_q$.

Let $a'=\langle\mathbf{a},\mathbf{v}\rangle$.
Thus,
\[
\left|
\left\{
\operatorname{Tr}(b_i\pi(z)-a' z)
:z\in\mathbb{F}_q
\right\}
\right|=1,
\]
contradicting the second condition of
Definition \ref{def:diff-split}.

Therefore,
\[
|W_{\phi'}(\mathbf{a},\mathbf{b})|<q^s.
\]
\qed    
\end{proof}

\begin{remark}
If symbol permutations are applied to fewer
than $s$ output coordinates of a linear AONT
$\phi$, then $W_{\max}(\phi')=q^s$.
Indeed, any unchanged output coordinate is
a linear function of the input, so we can
choose input and output masks for which
every term in the Walsh transform equals $1$.
Thus, permuting all $s$ output coordinates
is necessary to reduce the maximum Walsh
magnitude.
\end{remark}

We next show that applying difference-splitting permutations to all $s$ output columns will reduce the $W_{max}$ of a linear AONT.

\begin{proposition}\label{prop:all-column-permutation-walsh}
Let $\phi:\mathbb{F}_q^s\to\mathbb{F}_q^s$ be a
linear AONT, and let $\phi'$ be obtained by applying
difference-splitting permutations
$\pi_1,\ldots,\pi_s$ to all $s$ output coordinates
of $\phi$. Then
\[
W_{\max}(\phi')<q^s.
\]
\end{proposition}
\begin{proof}
Suppose, for a contradiction, that
$W_{\max}(\phi')=q^s$. Then there exist
$\mathbf{a}\in\mathbb{F}_q^s$ and
nonzero $\mathbf{b}\in\mathbb{F}_q^s$
such that
\[
|W_{\phi'}(\mathbf{a},\mathbf{b})|=q^s.
\]

Since $|W_{\phi'}(\mathbf{a},\mathbf{b})|=q^s$, there exists a constant $c\in\mathbb{F}_p$ such that
\[
\operatorname{Tr}
\left(
\sum_{j=1}^s b_j\pi_j(\phi(\mathbf{x})_j)
-\langle\mathbf{a},\mathbf{x}\rangle
\right)=c
\]
for every $\mathbf{x}\in\mathbb{F}_q^s$.

Since $\mathbf{b}\neq\mathbf{0}$, there exists
a coordinate $i$ with $b_i\neq0$. As $\phi$
is bijective, we can choose
$\mathbf{v}\in\mathbb{F}_q^s$ such that
$\phi(\mathbf{v})=\mathbf{e}_i$.

By linearity, $\phi(z\mathbf{v})=z\mathbf{e}_i$
for every $z\in\mathbb{F}_q$. Substituting
$\mathbf{x}=z\mathbf{v}$ into the
constant-phase condition gives
\[
\operatorname{Tr}
\left(
b_i\pi_i(z)
+\sum_{j\neq i}b_j\pi_j(0)
-z\langle\mathbf{a},\mathbf{v}\rangle
\right)=c
\]

Since the terms involving $\pi_j(0)$ are constant
with respect to $z$, we can move them to the
right-hand side. Writing
$a'=\langle\mathbf{a},\mathbf{v}\rangle$ and
\[
c'=c-\operatorname{Tr}
\left(\sum_{j\neq i}b_j\pi_j(0)\right)
\]
we obtain
\[
\operatorname{Tr}(b_i\pi_i(z)-a'z)=c'
\]
for every $z\in\mathbb{F}_q$.

Consequently,
\[
\left|
\left\{
\operatorname{Tr}(b_i\pi_i(z)-a'z)
:z\in\mathbb{F}_q
\right\}
\right|=1.
\]
This contradicts the second condition of
Definition \ref{def:diff-split}, since
$b_i\neq0$ and $\pi_i$ is difference-splitting.

Therefore,
\[
W_{\max}(\phi')<q^s.
\]
\qed
\end{proof}

We next establish a lower bound on the $W_{max}$ when permutations are applied only to the output coordinates of a linear AONT.

\begin{proposition}\label{prop:output-lower-bound-walsh}
Let $\phi:\mathbb{F}_q^s\to\mathbb{F}_q^s$ be a
linear AONT, and let $\phi'$ be obtained by applying
symbol permutations independently to its output
coordinates. Then
\[
W_{\max}(\phi')\geq\frac{q^s}{\sqrt{q-1}}
\]
\end{proposition}
\begin{proof}
Choose an output coordinate $i$ and let
$\mathbf{b}=\mathbf{e}_i$. Since $\phi$ is
linear, its $i$-th output coordinate is a
linear function of $\mathbf{x}$. Thus, for
every $t\in\mathbb{F}_q$, there exists an
input mask $\mathbf{a}_t$ such that
\[
\langle\mathbf{a}_t,\mathbf{x}\rangle
=t\phi(\mathbf{x})_i
\]
for every $\mathbf{x}\in\mathbb{F}_q^s$.

Let $\pi_i$ denote the permutation applied
to coordinate $i$. The corresponding Walsh
coefficient is
\[
W_{\phi'}(\mathbf{a}_t,\mathbf{e}_i)
=\sum_{\mathbf{x}\in\mathbb{F}_q^s}
\omega^{\operatorname{Tr}(
\pi_i(\phi(\mathbf{x})_i)-t\phi(\mathbf{x})_i)}
\]
Since $\phi$ is bijective, each value
$z\in\mathbb{F}_q$ occurs exactly
$q^{s-1}$ times in its $i$-th output
coordinate. Thus, setting
$z=\phi(\mathbf{x})_i$, we obtain
\[
W_{\phi'}(\mathbf{a}_t,\mathbf{e}_i)
=q^{s-1}\sum_{z\in\mathbb{F}_q}
\omega^{\operatorname{Tr}(\pi_i(z)-tz)}
\]

We now show that at least one of these
Walsh coefficients must have sufficiently
large magnitude. Applying Parseval's identity (Lemma \ref{lem:Parseval})
to $\pi_i$, with $s=1$ and output mask $b=1$,
gives
\[
\sum_{t\in\mathbb{F}_q}
\left|\sum_{z\in\mathbb{F}_q}
\omega^{\operatorname{Tr}(\pi_i(z)-tz)}
\right|^2=q^2
\]
Since $\pi_i$ is a permutation, the term
corresponding to $t=0$ is zero. Thus, the
remaining $q-1$ terms have squared magnitudes
summing to $q^2$, so by the averaging form of pigeonhole principle there exists some $t \neq 0$ such that
\[
\left|\sum_{z\in\mathbb{F}_q}
\omega^{\operatorname{Tr}(\pi_i(z)-tz)}
\right|^2
\geq\frac{q^2}{q-1}
\]

Taking square roots and using our earlier
expression for the Walsh coefficient, we obtain
\[
\begin{aligned}
W_{\max}(\phi')
&\geq |W_{\phi'}(\mathbf{a}_t,\mathbf{e}_i)|\\
&=q^{s-1}\left|\sum_{z\in\mathbb{F}_q}
\omega^{\operatorname{Tr}(\pi_i(z)-tz)}\right|\\
&\geq q^{s-1}\frac{q}{\sqrt{q-1}}\\
&=\frac{q^s}{\sqrt{q-1}}
\end{aligned}
\]
\qed
\end{proof}

We now consider the case where symbol permutations are applied to the input coordinates of a linear AONT. We show that this is equivalent, in terms of differential uniformity and maximum Walsh magnitude, to applying the same permutations to the output coordinates of the identity mapping. This allows us to then extend the previous results for output coordinate permutations to the input side.

\begin{proposition}\label{prop:input-side-equivalence}
Suppose
\[
\phi:\mathbb{F}_q^s\to\mathbb{F}_q^s
\]
is a linear AONT, and $\phi'$ is obtained by applying symbol permutations independently to the input coordinates of $\phi$.

Then the differential uniformity and maximum Walsh magnitude of $\phi'$ is equal to that of applying the same permutations to the output coordinates of the identity mapping over $\mathbb{F}_q^s$.
\end{proposition}

\begin{proof}
Let $T:\mathbb{F}_q^s\to\mathbb{F}_q^s$ represent the permutations applied to the input coordinates, defined by
\[
T(x_1,\ldots,x_s)
=
(\pi_1(x_1),\ldots,\pi_s(x_s)),
\]
where each $\pi_i$ is a permutation over $\mathbb{F}_q$. For any coordinate on which no permutation is applied, we take $\pi_i$ to be the identity permutation.

Applying these permutations to the inputs of $\phi$ gives
\[
\phi'(x)=\phi(T(x))
\]

It is known that affine equivalence preserves both
differential uniformity and nonlinearity \cite{bk-pnfandcrypto}. Thus, composing a mapping with an invertible linear transformation on the output side preserves its differential uniformity and maximum Walsh magnitude. 

Since $\phi$ is an invertible linear mapping,
\[
\Delta_{\phi'}=\Delta_T
\qquad\text{and}\qquad
W_{\max}(\phi')=W_{\max}(T)
\]

Now let $I$ be the identity mapping over $\mathbb{F}_q^s$. Observe that $T$ can also be obtained by applying the same symbol permutations to the output coordinates of $I$.

Therefore, both the differential uniformity and maximum Walsh magnitude of $\phi'$ are equal to those obtained by applying the same permutations to the output coordinates of the identity mapping.
\qed
\end{proof}

Notice that the preceding Propositions
\ref{Prop:diff_split-reduces},
\ref{Prop:single_col_limit},
\ref{prop:apply-on-all-columns},
\ref{prop:all-column-permutation} and
\ref{prop:output-lower-bound},
for differential uniformity and the corresponding Walsh Propositions
\ref{Prop:diff_split-reduces-walsh},
\ref{prop:all-column-permutation-walsh} and
\ref{prop:output-lower-bound-walsh},
only require that the underlying mapping be
linear and bijective, rather than specifically
an AONT. Since the identity mapping satisfies
these conditions, the previous results also
apply to $T$, and hence to input coordinate
permutations of $\phi$.

In particular, applying difference-splitting permutations to fewer than $s$ input coordinates leaves both $\Delta_{\phi'}$ and $W_{max}(\phi')$ equal to $q^s$, while applying them to all $s$ input coordinates reduces both below $q^s$. Furthermore, applying permutations only to the input coordinates cannot reduce the differential uniformity below $q^{s-1}$, and the $W_{max}$ below $(q^s)/(\sqrt{q-1})$.

This motivates applying permutations to both the input and output coordinates simultaneously, with the aim of achieving further reductions in differential uniformity and maximum Walsh magnitude.

By brute-force computation, we confirm that there are no difference-splitting permutations in $\mathbb{F}_3$ (the field, which coincides with $\mathbb{Z}_3$) or $\mathbb{F}_4$ (the field with 4 elements defined by $x^2+x+1$, not $\mathbb{Z}_4$). Thus any AONT constructed over alphabets of size $q= 3$ or $q=4$ using this method will not exhibit cryptographic non-linearity. 


Next we present three example constructions over fields where difference-splitting permutations exist for $q\geq 5$. In these examples, we construct non-linear AONTs by applying column-wise difference-splitting permutations on all columns. We report the resulting $\Delta_F$-uniformity and maximum Walsh magnitude obtained along with the construction. All the $W_{max}$ values reported below have been rounded up to the nearest integer and all permutations are permuting the field elements directly.

\begin{example}\label{ex:t25aont}
Consider the linear $(t,2,5)$-AONT generated by the matrix 
\setlength{\arraycolsep}{5pt}
\[
M=
\begin{pmatrix}
3 & 1\\
2 & 3 
\end{pmatrix}
\]
over $\mathbb{F}_5$.
Let $\pi_1$ be 
\[
\pi_1(0) = 1, \pi_1(1) = 0,
\pi_1(2) = 2, \pi_1(3) = 4, \text{ and } \pi_1(4) = 3.
\]
and let $\pi_2$ be 
\[
\pi_2(0) = 4, \pi_2(1) = 2,
\pi_2(2) = 3, \pi_2(3) = 0, \text{ and } \pi_2(4) = 1.
\]
Applying the symbol permutations $\pi_1$ on the input columns and $\pi_2$ on the output columns produces a cryptographically non-linear AONT with the differential $\Delta_F$-uniformity reduced from $25$ to $4$, and the maximum Walsh magnitude reduced from $25$ to $12$.
\end{example}

\begin{example} \label{ex:t37aont}
Consider the linear $(t,3,7)$-AONT generated by the matrix 
\setlength{\arraycolsep}{5pt}
\[
M=
\begin{pmatrix}
5 & 4 & 1\\
2 & 5 & 4\\
3 & 2 & 5
\end{pmatrix}
\]
over $\mathbb{F}_7$.
Let $\pi_1$ be 
\[
\pi_1(0) = 2,\quad \pi_1(1) = 0,
\pi_1(2) = 3,\quad \pi_1(3) = 6,
\]
\[
\pi_1(4) = 1,\quad \pi_1(5) = 4, 
\text{ and }\quad \pi_1(6) = 5.
\]
and let $\pi_2$ be 
\[
\pi_2(0) = 3,\quad \pi_2(1) = 0,
\pi_2(2) = 5,\quad \pi_2(3) = 1,
\]
\[
\pi_2(4) = 2,\quad \pi_2(5) = 4, 
\text{ and }\quad \pi_2(6) = 6.
\]
Applying the symbol permutations $\pi_1$ on the input columns and $\pi_2$ on the output columns produces a cryptographically non-linear AONT with the differential $\Delta_F$-uniformity reduced from $343$ to $10$ and the maximum Walsh magnitude reduced from $343$ to $63$.
\end{example}

\begin{example}\label{ex:t38aont}
Consider the linear $(t,3,8)$-AONT generated by the matrix 
\setlength{\arraycolsep}{5pt}
\[
M=
\begin{pmatrix}
x^2+1 & 1 & x^2\\
x^2 & x^2+x+1 & x^2+1\\
x+1 & x & x^2+x
\end{pmatrix}
\]
over $\mathbb{F}_8$ using $x^3+x+1$ as the irreducible polynomial. In this field we work with the following symbols:
\[
0,1,x,x^2,x+1,x^2+x,x^2+x+1,x^2+1
\]
Let $\pi_1$ be
\[
\begin{aligned}
\pi_1(0)&=0, & \pi_1(1)&=x+1, &\pi_1(x)&=x^2+1, &\pi_1(x^2)&=x,\\
\pi_1(x+1)&=1, & \pi_1(x^2+x)&=x^2, &\pi_1(x^2+x+1)&=x^2+x,&\pi_1(x^2+1)&=x^2+x+1.
\end{aligned}
\]
and let $\pi_2$ be
\[
\begin{aligned}
\pi_2(0)&=x^2+x, & \pi_2(1)&=0,&\pi_2(x)&=x, & \pi_2(x^2)&=x+1,\\
\pi_2(x+1)&=x^2+1, & \pi_2(x^2+x)&=1,&\pi_2(x^2+x+1)&=x^2, &\pi_2(x^2+1)&=x^2+x+1.
\end{aligned}
\]
Applying the symbol permutations $\pi_1$ on the input columns and $\pi_2$ on the output columns, produces a cryptographically non-linear AONT with the differential $\Delta_F$-uniformity reduced from $512$ to $12$ and the maximum Walsh magnitude reduced from $512$ to $128$.

Worked calculations of differential uniformity and Walsh magnitude for Examples \ref{ex:t25aont} and \ref{ex:t38aont} are provided in Appendix~\ref{sec:Appendix}.

\end{example}

\section{Randomized Search and Experiment}
\label{sec:Randomized-Search}

In this section we use computational experiments to further study how coordinate-wise permutations affect the differential uniformity and maximum Walsh magnitude of linear AONTs. We use Cauchy matrices to construct the linear AONTs \cite{des-allornothing}, as they provide the strongest possible linear AONT, acting simultaneously as a $(t,s,q)$-AONT for all $t$ such that $1 \leq t \leq s$. Our goal is twofold: to find AONTs with small $\Delta_F$ and $W_{max}$ through randomized search, and to estimate the mean and standard deviation of these measures resulting from random permutations. The results offer insight into the $\Delta$ and $W_{max}$ values achievable through coordinate-wise permutations.

The Rust implementations of the algorithms in this section are available here: 

\url{https://github.com/SPyCy-Lab/AONT-Search}.

\subsection{Randomized Search}

Computing the exact differential uniformity of an AONT requires $O(q^{2s})$ operations and computing the maximum Walsh magnitude using a fast Fourier transform requires $O(sq^{2s}\log q)$
operations. These computational costs make it expensive to evaluate a large number of candidates. To make the search feasible, we adopt a two-stage approach in which promising candidates are first identified through a less expensive evaluation, followed by an exact computation of their differential uniformity or maximum Walsh magnitude.

We begin with a linear AONT and generate new candidates by independently applying $s$ input permutations and $s$ output permutations over the underlying finite field. Each permutation is selected uniformly at random from the $q!$ possible permutations of the field elements. In our Rust implementation, this is achieved using the \texttt{shuffle()} function from the \texttt{rand} crate, which implements the Fisher–Yates shuffle algorithm.

For differential uniformity, rather than computing the exact value for every candidate, we initially evaluate each one over a fixed subset of approximately $10^4$ nonzero input differences. The largest differential count observed over this subset provides a lower bound on the candidate's differential uniformity. We repeat this process for approximately $10^5$ randomly generated candidates, recording the permutations whenever a smaller lower bound is found.
For the Walsh transform, we follow the same randomized search procedure, but due to the increased complexity we evaluate approximately $10^3$ candidates using a subset of $10^4$ output masks for each candidate. The largest Walsh magnitude observed over this subset provides a lower bound on the candidate's maximum Walsh magnitude. As before, we record the permutations whenever an improvement is found.

Once the search is complete, we select the candidate with the smallest observed lower bound and compute its exact differential uniformity or maximum Walsh magnitude. The lowest $\Delta_F$ and lowest $W_{max}$ found by the search for multiple parameters are reported in tables \ref{tab:randomized-search-results-delta} and \ref{tab:randomized-search-results-Wmax}, respectively.

\subsection{Statistical Experiment}

To study the typical effect of coordinate-wise permutations, we also
conduct a separate experiment in which randomly generated candidates
are evaluated without the preliminary sampling stage.

For each parameter set, we independently generate 100 sets of input
and output permutations and compute the exact differential uniformity and maximum Walsh magnitude of each resulting AONT. From these values, we calculate the sample mean, median and standard deviation.

Unlike the randomized search, which aims to identify candidates with
particularly small differential uniformity, this experiment provides
an estimate of the distribution of these measures under
uniformly random coordinate-wise permutations.

\begin{table}[htbp]
    \centering
    \caption{Results of the randomized search and statistical experiment for differential uniformity. The lowest $\Delta$ is computed using the screened search, while mean, median, and standard deviation are based on $100$ exact samples.}
    \label{tab:randomized-search-results-delta}
    \begin{tabular}{|c|c|c|c|c|}
        \hline
        $(t,s,q)$ &
        Lowest $\Delta$ found &
        Mean  &
        Median &
        Std. dev. \\
        \hline
         $(t,4,11)$&23 & 104.71&98.5 &30.98 \\
        \hline
         $(t,5,11)$&84 & 423.18&366.5 & 196.98 \\
        \hline
         $(t,4,13)$&27 & 107.87&97 & 41.20\\
        \hline
         $(t,5,13)$&111 &461.31 &400.5 &187.91\\
        \hline
         $(t,6,13)$ &541 &1963.05 &1784 &816.56\\
        \hline
         $(t,4,17)$&33 &94.58 &91 &27.93\\
        \hline
        $(t,4,31)$&50 &88.33 &85.5 &19.41\\
        \hline
        $(t,4,43)$&47 &79.57 &75 &17.18\\
        \hline
        $(t,4,47)$&53 &78.45 &76 &15.76\\
        \hline
        $(t,2,257)$&11 &12.06 &12 &0.51\\
        \hline
    \end{tabular}
\end{table}

\begin{table}[hbtp]
    \centering
    \caption{Results of the randomized search and statistical experiment for maximum Walsh magnitude. The lowest $W_{max}$ is computed using the screened search, while mean, median, and standard deviation are based on $100$ exact samples. The lower bound of $q^{s/2}$ and all $W_{max}$ results shown in the table have been rounded up to the nearest integer.}
    \label{tab:randomized-search-results-Wmax}
    \begin{tabular}{|c|c|c|c|c|c|}
        \hline
        $(t,s,q)$ &
        $q^{s/2}$&        
        Lowest $W_{max}$ found &
        Mean  &
        Median &
        Std. dev. \\
        \hline
         $(t,4,11)$&121 &813 & 1684.06 & 1683 & 332.68 \\
        \hline
         $(t,5,11)$&401 &6867 & 12109.91 & 11547 & 2584.32 \\
        \hline
         $(t,4,13)$&169 &1384 & 2406.01 & 2315 & 525.46\\
        \hline
         $(t,5,13)$&609 &11839 &19538.97 & 19337 & 3444.66\\
        \hline
         $(t,6,13)$&2197 &101380 &162717.81 & 157374 & 32362.68\\
        \hline
         $(t,4,17)$&289 &2607 & 4262.63 & 4171 & 806.18\\
        \hline
        $(t,4,31)$&961 &11352 & 15221.81 & 15171 & 1896.57\\
        \hline
        $(t,4,43)$&1849 &21836 & 29802.01 & 29358 & 4642.7\\
        \hline
        $(t,4,47)$&2209 &27063 & 35199.72 & 34403 &4933.58\\
        \hline
        $(t,2,257)$&257 &1139 &1206.53 &1199 &37.4\\
        \hline
    \end{tabular}
\end{table}

The computational results show that coordinate-wise permutations can produce AONTs with relatively small differential uniformity and $W_{max}$, although the resulting values do vary considerably depending on the permutations chosen. The randomized search consistently identified values substantially below the averages observed in the statistical experiment. These findings suggest that carefully chosen permutations can significantly reduce these measures and motivate further investigation into the mathematical structure of these permutations. It should be noted that these minima are conservative and they are not comparable across parameter sets because the same number of the differences and the candidates are used in the prescreening step for all parameter sets, with search-spaces sizes from $11^4 = 14641$ to $47^4 = 4,879,681$, and only the single best candidate from the prescreening is evaluated exactly.

\section{Conclusion}\label{sec:conclusion}

In this paper, we have initiated the study of non-linear unconditionally secure AONTs. Our contributions are threefold:

\begin{enumerate}
    \item \textbf{Enumeration and non-existence result:} we exhaustively enumerated all general $(2,3,3)$-AONTs, finding 192 linear and 4992 non-additive instances; by Corollary \ref{Cor:AONT-Inverse}, this result also yields the complete enumeration of $(1,3,3)$-AONTs. Considering limits in our computational resources, these two are the only parameter-sets which we can enumerate exhaustively. Notably, computational analysis revealed that all enumerated non-additive $(t,3,3)$-AONTs are affine: they attain maximum differential $\Delta_F$-uniformity and maximum Walsh magnitude. Thus, although these mappings are non-additive,  they are affine and not cryptographically non-linear. This observation suggests that the small parameter set $(t,3,3)$ is insufficient to obtain cryptographically meaningful non-linearity, and motivates the study of larger parameter sets. 
    \item \textbf{Construction and Characterization:} we presented a general construction (Construction \ref{Con:nonlinear-construction}) for obtaining non-additive and non-affine AONTs from linear ones, by means of column-wise symbol permutations. We then identified the precise mechanism by which such permutations reduce $\Delta_F$-uniformity and $W_{max}$ through the notion of \emph{difference-splitting} permutations (see Definition \ref{def:diff-split}), which we showed are exactly the non-affine permutations of $\mathbb{F}_p$, for prime values of $p$.  Proposition \ref{Prop:diff_split-reduces} shows that a single difference-splitting permutation applied to one output coordinate strictly reduces the differential count $\delta_\phi(a,\phi(a))$ for every nonzero $a$ whose image $\phi(a)$ is nonzero in that coordinate. Proposition \ref{Prop:single_col_limit}, conversely, establishes a fundamental limitation: when the relevant output coordinate of $\phi(a)$ is $0$, the differential count at $(a,\phi(a))$ is preserved under any permutation of that column, and Proposition \ref{prop:apply-on-all-columns} extends this result to any construction that permutes fewer than $s$ output coordinates.
    Similarly, Proposition \ref{Prop:diff_split-reduces-walsh} shows that applying a difference-splitting permutation on a single column reduces the Walsh magnitude for every $a$ and any non-zero $b$ supported on that coordinate. Proposition \ref{prop:all-column-permutation-walsh} shows that applying difference-splitting permutations on all output columns strictly reduces the maximum Walsh magnitude. Lastly, Proposition \ref{prop:output-lower-bound-walsh} presents a lower bound on the maximum Walsh magnitude achievable with output side permutations. Together these six results motivate applying difference-splitting permutations \emph{independently to all output coordinates and input coordinates}, which is exactly the setting of Proposition \ref{prop:input-side-equivalence}: under this construction, $\Delta_{\phi'} < q^s$ and $W_{max} < q^s$ for the resulting AONT. 
    
    Since $\mathbb{F}_3$ and $\mathbb{F}_4$ admit no difference-splitting permutations, no constructions of this form can yield cryptographically non-linear AONTs over these alphabets, which explains why our enumeration of $(t,3,3)$-AONTs found only affine non-additive instances. Examples \ref{ex:t25aont}--\ref{ex:t38aont} illustrate the constructions for $q = 5, 7, 8$, where difference-splitting permutations exist and substantial reductions in both differential $\Delta_F$-uniformity and maximum Walsh magnitude are achieved. 
    
    \item \textbf{Computational results:} Using randomized search, we found two-sided permutations achieving $\Delta_F$ values far below the one-sided bound of $q^{s-1}$. Our statistical experiment further suggests that permutations yielding relatively small $\Delta_F$ and $W_{max}$ are common rather than rare.
\end{enumerate}




\paragraph{Open problems.} Several directions remain open.

First, can one characterize \emph{quantitatively} which combinations of 
column permutations yield the lowest differential $\Delta_F$-uniformity for a given
linear AONT? More precisely, given a linear $(t,s,q)$-AONT $\phi$, what is the resulting $\Delta_{\phi'}$ as a function of $s$ and $q$, and what is the minimum $\Delta_{\phi'}$ achievable
over all such independent column-wise permutations on input-side and output-side?

Second, is there a non-additive AONT that is not
obtainable from any linear AONT via independent column-wise symbol permutations? Proposition \ref{Prop:allt33attainable} answers this negatively for $(t,3,3)$-AONTs; the question is whether the result extends to larger parameter sets.

Finally, what is the relationship between cryptographically non-linear AONTs and known classes
of cryptographically optimal functions? In particular, the
Chabaud-Vaudenay \cite{cv-bound} link between $\Delta_F$ and the Walsh spectrum suggests
that the achievable $W_{\max}$ of an AONT obtained from
column permutations is bounded below by a
quantity involving $\Delta_{\phi'}$.


 \section*{Acknowledgments}
 
 The authors would like to thank Alfred Menezes, Ian Goldberg, and Daniel Ibanescu, who on three separate occasions raised questions about the existence of non-linear AONTs, which eventually led to this work.

\bibliographystyle{splncs04}
\bibliography{nlaont}

\section{Appendix}\label{sec:Appendix}
\subsection{Worked Examples of Differential Uniformity and Walsh Magnitude}

We illustrate the computation of differential uniformity and
Walsh magnitude for the constructions in Examples 1 and 3.
Rather than evaluating every possible input difference and
Walsh mask, we consider particular pairs $(a,b)$ that attain
the maximum values reported in Section~\ref{sec:measuring-non-linearity}.
For each construction, we first demonstrate how the
differential equation is evaluated and identify all inputs
satisfying it for the selected pair. We then compute a
Walsh coefficient by grouping the contributions of individual
inputs according to their character values.

These calculations illustrate how the two non-linearity
measures are obtained. Establishing that the resulting
values are global maxima requires considering all admissible
pairs $(a,b)$. These exhaustive computations were performed independently in Rust; the code is available at \url{https://github.com/SPyCy-Lab/AONT-Search}.

Throughout the calculation, the AONT mapping is represented as such
\[
F(u)=\pi_2\bigl(\pi_1(u)M^{-1}\bigr),
\]
where the permutations act coordinatewise and vectors are
treated as row vectors.

\paragraph{Example 1: A construction over $\mathbb F_5$.}

Consider the matrix and permutations from Example 1,
\[
M=\begin{pmatrix}3&1\\2&3\end{pmatrix},
\qquad
M^{-1}=\begin{pmatrix}4&2\\4&4\end{pmatrix},
\]
with
\[
\pi_1=(1,0,2,4,3),
\qquad
\pi_2=(4,2,3,0,1),
\]
where each permutation is specified by its images on
$0,1,2,3,4$.

To illustrate the differential calculation, consider
\[
a=(0,1),\qquad b=(1,1).
\]
We seek the inputs $u\in\mathbb F_5^2$ satisfying
\[
F(u+a)-F(u)=b.
\]

For instance, taking $u=(0,0)$, we obtain
\[
\begin{aligned}
\pi_1(u)&=(1,1),\\
\pi_1(u)M^{-1}&=(3,1),\\
F(u)&=\pi_2(3,1)=(0,2).
\end{aligned}
\]
Similarly, since $u+a=(0,1)$,
\[
\begin{aligned}
\pi_1(u+a)&=(1,0),\\
\pi_1(u+a)M^{-1}&=(4,2),\\
F(u+a)&=\pi_2(4,2)=(1,3).
\end{aligned}
\]
Consequently,
\[
F(u+a)-F(u)=(1,3)-(0,2)=(1,1)=b.
\]

Repeating this calculation for every
$u\in\mathbb F_5^2$, we find that the complete
set of solutions is
\[
\{(0,0),(0,2),(1,1),(1,3)\}.
\]
Thus, for this particular pair,
\[
\delta((0,1),(1,1))=4.
\]

Exhaustive evaluation confirms that no other
nonzero input difference and output difference produce
a larger count, so we obtain $\Delta_F=4$.

We next illustrate the Walsh calculation using
\[
a=(2,1),\qquad b=(0,1).
\]
Here the absolute trace is the identity, since
the underlying field is $\mathbb F_5$.

For example, when $u=(0,0)$, we have
$F(u)=(0,2)$, and therefore
\[
\begin{aligned}
\langle b,F(u)\rangle-\langle a,u\rangle
&=(0,1)\cdot(0,2)-(2,1)\cdot(0,0)\\
&=2.
\end{aligned}
\]
Hence this input contributes $\omega^2$ to
$W_F(a,b)$.

Rather than listing the contributions of all
$25$ inputs individually, we group them according
to the value of the exponent. The number of inputs
giving exponents $0,1,2,3,4$, respectively, are
\[
(0,6,12,3,4).
\]
It follows that
\[
\begin{aligned}
W_F(a,b)
&=6\omega+12\omega^2+3\omega^3+4\omega^4\\
&\approx -9.045085+7.192180i,
\end{aligned}
\]
and consequently
\[
|W_F(a,b)|\approx 11.555995.
\]
An exhaustive evaluation over all admissible Walsh
masks confirms that
\[
W_{\max}(F)\approx 11.555995.
\]
This value is reported as $12$ after rounding up to the nearest integer.

\paragraph{Example 3: A construction over $\mathbb F_8$.}

We now consider Example 3 to illustrate the corresponding
calculations over an extension field, where both field
multiplication and the absolute trace must be taken into
account.

Recall that
\[
\mathbb F_8=\mathbb F_2[x]/(x^3+x+1),
\]
so that $x^3=x+1$.

For the matrix $M$ from Example 3, its inverse is
\[
M^{-1}=
\begin{pmatrix}
x^2+x+1 & x^2+x+1 & x\\
x^2+1   & x         & x^2+1\\
x^2+1   & 1         & 1
\end{pmatrix}.
\]
The input permutation $\pi_1$ is given by
\[
\begin{aligned}
\pi_1(0)&=0, &
\pi_1(1)&=x+1,\\
\pi_1(x)&=x^2+1, &
\pi_1(x^2)&=x,\\
\pi_1(x+1)&=1, &
\pi_1(x^2+x)&=x^2,\\
\pi_1(x^2+x+1)&=x^2+x, &
\pi_1(x^2+1)&=x^2+x+1.
\end{aligned}
\]
Similarly, the output permutation $\pi_2$ is
\[
\begin{aligned}
\pi_2(0)&=x^2+x, &
\pi_2(1)&=0,\\
\pi_2(x)&=x, &
\pi_2(x^2)&=x+1,\\
\pi_2(x+1)&=x^2+1, &
\pi_2(x^2+x)&=1,\\
\pi_2(x^2+x+1)&=x^2, &
\pi_2(x^2+1)&=x^2+x+1.
\end{aligned}
\]
Both permutations are applied coordinatewise.

To illustrate the differential calculation, consider
\[
a=(0,x^2,1),
\qquad
b=(x^2+x+1,0,x^2+x).
\]
We seek the inputs $u\in\mathbb F_8^3$ satisfying
\[
F(u+a)-F(u)=b.
\]

For instance, taking
\[
u=(0,x,x^2+x),
\]
we first apply the input permutation:
\[
\pi_1(u)=(0,x^2+1,x^2).
\]
Multiplication by $M^{-1}$ gives
\[
\pi_1(u)M^{-1}=(x^2+1,x^2+1,x+1).
\]
For example, the first coordinate is obtained by
evaluating
\[
\begin{aligned}
&0(x^2+x+1)
 +(x^2+1)(x^2+1)
 +x^2(x^2+1)\\
&\qquad=x^2+1,
\end{aligned}
\]
where multiplication is performed modulo
$x^3+x+1$.

Applying the output permutation then gives
\[
F(u)=(x^2+x+1,x^2+x+1,x^2+1).
\]

Similarly,
\[
u+a=(0,x^2+x,x^2+x+1).
\]
Applying the input permutation, we obtain
\[
\pi_1(u+a)=(0,x^2,x^2+x),
\]
and consequently
\[
\pi_1(u+a)M^{-1}=(1,x^2+1,x^2).
\]
Applying $\pi_2$ gives
\[
F(u+a)=(0,x^2+x+1,x+1).
\]

Since $\mathbb F_8$ has characteristic two,
subtraction coincides with addition. Hence
\[
\begin{aligned}
F(u+a)-F(u)
&=(x^2+x+1,0,x^2+x)\\
&=b.
\end{aligned}
\]

Repeating this calculation for all $512$ inputs,
we obtain the following complete set of solutions:
\[
\begin{aligned}
\mathcal S=\{&
(0,x,x^2+x),\\
&(0,x^2+x,x^2+x+1),\\
&(1,0,x+1),\\
&(1,x^2,x),\\
&(x,x+1,x^2),\\
&(x,x^2+x+1,x^2+1),\\
&(x+1,x+1,x^2),\\
&(x+1,x^2+x+1,x^2+1),\\
&(x^2+x,0,x+1),\\
&(x^2+x,x^2,x),\\
&(x^2+x+1,x,x^2+x),\\
&(x^2+x+1,x^2+x,x^2+x+1)
\}.
\end{aligned}
\]
Therefore, for this particular pair,
\[
\delta(a,b)=|\mathcal S|=12.
\]
Exhaustive evaluation over all nonzero input
differences and output differences confirms that
\[
\Delta_F=12.
\]

We next illustrate the Walsh calculation, where
the absolute trace determines the contribution of
each input to the character sum. Consider the masks
\[
a=(x^2+1,x^2+x+1,x^2+x),
\qquad
b=(0,0,x).
\]
For the input $u=(0,x,x^2+x)$ considered above,
we already calculated
\[
F(u)=(x^2+x+1,x^2+x+1,x^2+1).
\]

The output inner product is
\[
\begin{aligned}
\langle b,F(u)\rangle
&=x(x^2+1)\\
&=x^3+x\\
&=1,
\end{aligned}
\]
where we used $x^3=x+1$.
Likewise, the input inner product is
\[
\begin{aligned}
\langle a,u\rangle
&=(x^2+1)\cdot 0+(x^2+x+1)\cdot x+(x^2+x)(x^2+x)\\
&=0+(x^2+1)+x\\
&=x^2+x+1.
\end{aligned}
\]
Thus,
\[
\begin{aligned}
z
&=\langle b,F(u)\rangle-\langle a,u\rangle\\
&=1+(x^2+x+1) \qquad \text{(characteristic 2)}\\
&=x^2+x.
\end{aligned}
\]

Unlike the prime-field example, we must apply the
absolute trace to obtain an element of $\mathbb F_2$.
In this case,
\[
\operatorname{Tr}(z)=z+z^2+z^4.
\]
Using the defining polynomial, we calculate
\[
z^2=(x^2+x)^2=x,
\qquad
z^4=x^2.
\]
Consequently,
\[
\begin{aligned}
\operatorname{Tr}(z)
&=(x^2+x)+x+x^2\\
&=0.
\end{aligned}
\]
Since $\omega=e^{2\pi i/2}=-1$, this input
contributes
\[
\omega^{\operatorname{Tr}(z)}
=(-1)^0=1
\]
to $W_F(a,b)$.

Repeating this procedure for all $512$ inputs,
we find that the trace is zero for $320$ inputs
and one for the remaining $192$. Grouping the
corresponding character contributions gives
\[
\begin{aligned}
W_F(a,b)
&=320\omega^0+192\omega^1\\
&=320-192\\
&=128.
\end{aligned}
\]
Thus $|W_F(a,b)|=128$ for the selected masks.
Exhaustive evaluation over all admissible Walsh
masks confirms that
\[
W_{\max}(F)=128.
\]
These examples demonstrate how differential
uniformity is obtained by counting solutions to
the differential equation, whereas the Walsh
magnitude is obtained by summing additive
characters over the entire input space.
In the extension-field case, the absolute trace
determines the character contribution of each input.

\end{document}